\documentclass{IEEEtran}
\pdfoutput=1

\ifCLASSINFOpdf
\else
\fi
\usepackage{amsmath,amssymb,amsfonts,bm}
\usepackage{graphicx,subfig}
\usepackage{cite}
\usepackage{multirow}
\usepackage{float}
\usepackage{cuted}
\begin{document}
\title{Joint RFI Mitigation and Radar Echo Recovery for One-Bit UWB Radar}
\author{Tianyi~Zhang,~% <-this % stops a space
Jiaying~Ren,
Jian~Li,~\IEEEmembership{Fellow,~IEEE,}
Lam~H.~Nguyen
and~Petre~Stoica,~\IEEEmembership{Fellow,~IEEE,}
\thanks{This work was supported in part by the National Science Foundation under Grant 1704240 and in part by the Swedish Research Council under VR grants 2017-04610 and 2016-06079 (corresponding author: Jian Li).}
\thanks{T. Zhang, J. Ren, and J. Li are with the department od Electrical and Computer Engineering, University of Florida, Gainesville, FL 32611, USA (e-mail: tianyi.zhang@ufl.edu, jiaying.ren@ufl.edu, li@dsp.ufl.edu).}
\thanks{L.H.Nguyen is with the U.S. Army Research Laboratory, Adelphi, MD 20783 USA (e-mail: lam.h.nguyen2.civ@mail.mil).}
\thanks{P. Stoica is with the department of Information Technology, Uppsala University, P. O. Box 337, SE-751 05 Uppsala, Sweden (e-mail: ps@it.uu.se).}
}

\maketitle

\begin{abstract}
Radio frequency interference (RFI) mitigation and radar echo recovery are critically important for the proper functioning of ultra-wideband (UWB) radar systems using one-bit sampling techniques. We recently introduced a technique for one-bit UWB radar, which first uses a majorization-minimization method for RFI parameter estimation followed by a sparse method for radar echo recovery. However, this technique suffers from high computational complexity due to the need to estimate the parameters of each RFI source separately and iteratively. In this paper, we present a computationally efficient joint RFI mitigation and radar echo recovery framework to greatly reduce the computational cost. Specifically, we exploit the sparsity of RFI in the fast-frequency domain and the sparsity of radar echoes in the fast-time domain to design a one-bit weighted SPICE (SParse Iterative Covariance-based Estimation) based framework for the joint RFI mitigation and radar echo recovery of one-bit UWB radar. Both simulated and experimental results are presented to show that the proposed one-bit weighted SPICE framework can not only reduce the computational cost but also outperform the existing approach for decoupled RFI mitigation and radar echo recovery of one-bit UWB radar.
\end{abstract}


% \fi
\IEEEpeerreviewmaketitle

\section{Introduction}
\IEEEPARstart{U}{ltra}-wideband (UWB) radar systems have a wide range of applications including, for example, landmine and unexploded ordinance (UXO) detection using ground penetrating radar (GPR) \cite{CHND01}, hidden object imaging via foliage penetrating radar (FOPEN) radar \cite{XN01}, as well as human detection \cite{YLML06} and non-contact human vital sign monitoring\cite{SLL20}. However, due to the large bandwidth, an analog-to-digital converter (ADC) with a high-sampling rate is necessary at the UWB radar receiver. For example, for an impulse UWB radar system with a bandwidth larger than 10 GHz, the sampling rate of the ADC should be larger than 20 GHz. However, a radar system using such an ADC, especially with a high quantization precision, may be too expensive to be commercially viable. Even if a high-speed ADC with high-resolution quantization was available, it would greatly increase the cost and power consumption of the UWB radar system. To solve this problem, an ADC with low-resolution quantization can be attractive due to its low cost and low power consumption advantages, as well as its potential to achieve ultra-high sampling rates \cite{ZHLLW18,ZHB19}. For instance, the NVA6100 impulse radar system, a single-chip UWB radar from Novelda, utilizes the so-called Continuous Time Binary Value (CTBV) technique to achieve a very high sampling rate of 39 GHz and a 13-bit quantization precision \cite{Xethru,HWLLM07}. CTBV is an efficient one-bit sampling strategy, an extreme form of low-bit resolution sampling, which obtains its signed measurements via comparing the received signal to a known threshold. The threshold varies linearly with slow-time, i.e., from one pulse repetition interval (PRI) to another. High-precision data samples can then be obtained from the signed measurements (i.e., one-bit fast-time samples for each PRI) via a simple digital integration (DI) method \cite{HWLLM07} under the assumption that the targets are stationary during the short time interval of the sample collection process. Leveraging the CTBV technology, the NVA6100 system employs simple circuitry and has low power consumption. NVA6100 can be used for diverse applications, including vital sign monitoring \cite{SLL20}, through-wall imaging and object tracking \cite{Xethru}. We refer to the CTBV-based UWB radar system as the one-bit UWB radar in this paper.

One of the most significant problems of a UWB radar system is the presence of strong radio frequency interferences (RFIs) caused by many competing users within the ultra-wide frequency band of the radar. Typical RFI sources include FM radio transmitters, TV broadcast transmitters, cellular phones, and other radiation devices whose operating frequencies tend to overlap with those of the UWB radar systems \cite{KL95}. These RFI sources pose a significant hindrance to the proper operations of the UWB radar systems in terms of reduced signal-to-noise ratio (SNR) and degraded radar imaging quality.

RFI mitigation is a notoriously challenging problem since it is difficult to predict and model RFI signals accurately due to their dynamic range and diverse modulation schemes. We have recently developed a technique for one-bit UWB radar, which first estimates the RFI parameters \cite{ZRGL19,ZRLL20}, using 1bMMRELAX-1bBIC algorithm \cite{RZLS19} and then uses a sparse method for echo recovery. However, this technique suffers from high computational complexity due to the need of using 1bMMRELAX-1bBIC to estimate the parameters of each RFI source separately and iteratively.

In this paper, we present a computationally efficient joint RFI mitigation and radar echo recovery framework for one-bit UWB radar systems. The echo signals of an impulse UWB radar are commonly sparse in the fast-time domain due to the sparsity of strong targets present in the scene of interest. Similarly, the RFI sources also tend to be sparse but in the fast-frequency domain (see, e.g., Figure \ref{fig:spec_RFI_mea}). We exploit these properties and introduce a one-bit weighted SPICE framework to jointly mitigate the RFI and recover the radar echoes from signed measurements. Inspired by the original weighted SPICE framework devised for high-precision data sets\cite{SZL2014}, a weighted SPICE technique was presented in \cite{SLL20-2} for target parameter estimation for one-bit automotive radar systems. However, this method was designed for single-PRI data containing only radar echoes contaminated by noise. The method in \cite{SLL20-2} cannot be directly applied to our problem, i.e., radar echo recovery from multiple-PRI data sets containing severe RFI and noise. Our main contributions can be summarized as follows:

\begin{figure}[htbp]
\centering
\includegraphics[width=0.45\textwidth]{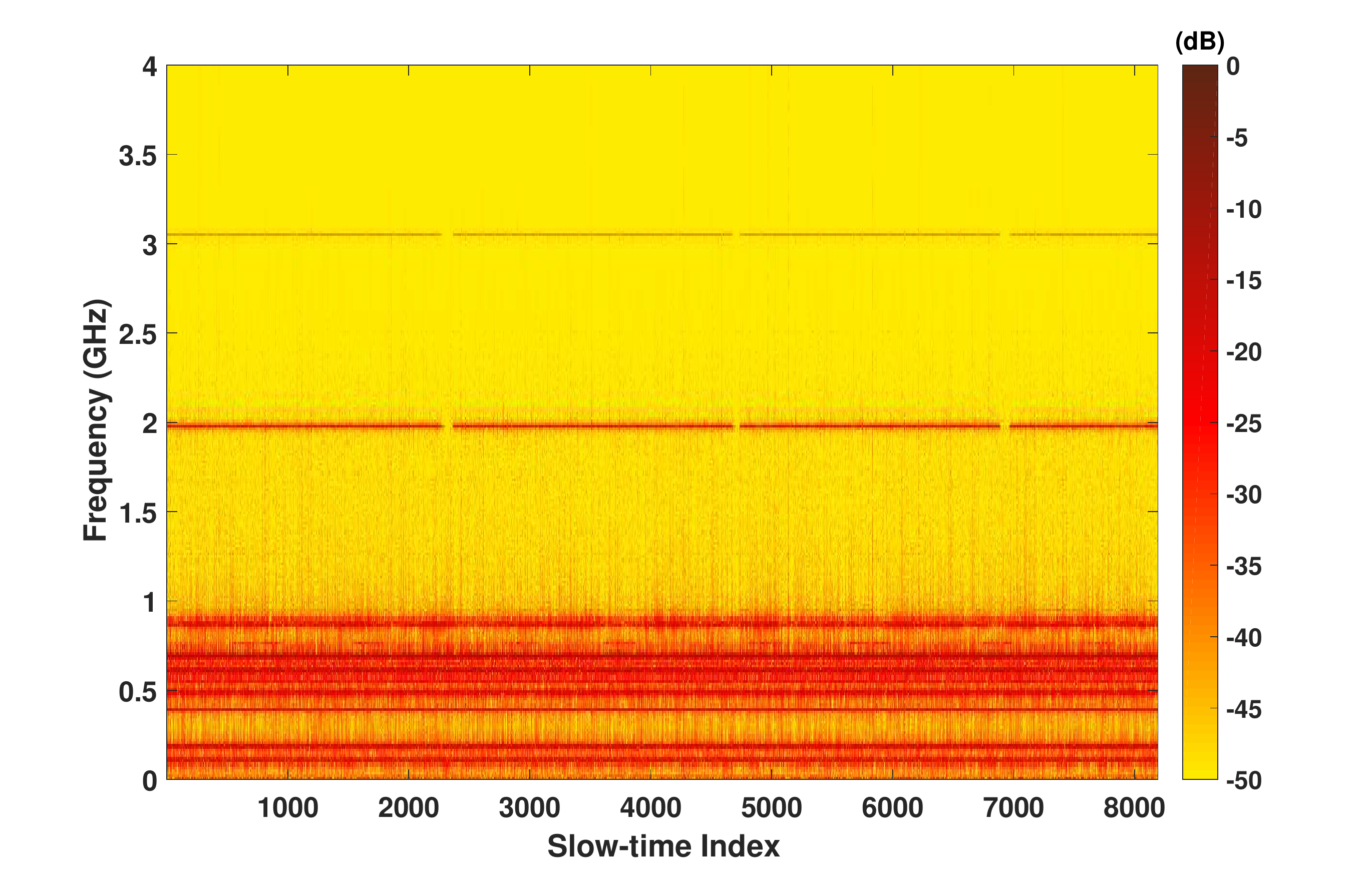}
\caption{Example of fast-time RFI spectrum vs. slow-time index, for the RFI-only data measured by an experimental ARL UWB radar receiver.}
\label{fig:spec_RFI_mea}
\end{figure}
1) We establish a proper data model for the signed measurements obtained by the one-bit UWB radar systems.

2) We jointly exploit the sparsity of radar echoes in the fast-time domain and the sparsity of the RFI sources in the fast-frequency domain.

3) We introduce a one-bit weighted SPICE based framework to jointly mitigate the RFI sources and recover the radar echoes from the signed measurements of the one-bit UWB radar systems. The one-bit weighted SPICE framework in \cite{SLL20-2} is modified herein to deal with multiple-PRI data sets and with UWB radar echoes contaminated by severe RFI and noise. The proposed computationally efficient joint RFI mitigation and radar echo recovery methodology reduces the computational cost of the existing technique in \cite{ZRGL19, ZRLL20} while providing improved performance of RFI mitigation and radar echo recovery.

4) Both simulated and measured RFI data sets are used in this paper to demonstrate the effectiveness of the one-bit weighted SPICE based framework for joint RFI mitigation and radar echo recovery. The impact of various weighting choices on the performance of the one-bit weighted SPICE is also discussed herein.

The rest of this paper is organized as follows. In Section \ref{sec:PF}, we formulate the RFI mitigation and radar echo recovery problem for a one-bit UWB radar system using the CTBV sampling technique.  Next in Section \ref{sec:c_spice}, we briefly review the weighted SPICE framework for high-precision measurements. In Section \ref{sec:1bSPICE}, we present the one-bit weighted SPICE framework for joint RFI mitigation and radar echo recovery for one-bit UWB radar. Finally, in Section \ref{sec:experiment}, we provide both simulated and experimental examples to demonstrate the improved performance and reduced computational cost of the proposed one-bit weighted SPICE framework.

\emph{Notation:} We denote vectors and matrices by boldface lower-case and upper-case letters, respectively. $(\cdot)^T$ and $(\cdot)^H$ denote the transpose and conjugate transpose operations, respectively. ${\bm x}_{n,:}$ and ${\bm x}_{:,m}$ denote the $n$-th row and $m$-th column of matrix ${\bm X}$, respectively. $x_{n,m}$ denotes the $(n,m)$-th element of the matrix ${\bm X}$, and $x_n$ denotes the $n$-th element of the vector ${\bm x}$. For a matrix or a vector, $\left\|\cdot\right\|_p$ denotes the $\ell_p$ norm, i.e., $\left\|{\bm X}\right\|_p = \left(\sum_{m=1}^M\sum_{n=1}^N |x_{n,m}|^p\right)^{1/p}$ or $\left\|{\bm x}\right\|_p = \left(\sum_{n=1}^N |x_n|^p\right)^{1/p}$. ${\bm 1}_N = [1,\dots,1]^T \in\mathbb{R}^{N\times 1}$ and ${\bm I}_N$ denotes the $N\times N$ identity matrix. $\odot$ denotes the element-wise product of matrices or vectors. The acronyms used in this paper are summarized in Table \ref{tab:acronyms}:
\begin{table}[htbp]
\centering
\begin{tabular}{|l|l|}
\hline
Acronyms & Complete Name\\
\hline
ADC         & Analog-to-Digital Converter \\
%DAC         & digital-to-analog converter \\
UWB         & Ultra-WideBand \\
CPI         & Coherent Processing Interval \\
PRI         & Pulse Repetition Interval \\
CTBV        & Continuous Time Binary Value \\
RFI         & Radio Frequency Interference \\
SPICE       & SParse Iterative Covariance-based Estimation \\
LIKES       & LIKelihood-based Estimation of Sparse parameters \\
IAA         & Iterative Adaptive Approach\\
MM          & Majorization-Minimization\\
DI          & Digital Integration \\
ARL         & Army Research Laboratory \\
BIC         & Bayesian Information Criterion \\
SINR        & Signal-to-Interference-plus-Noise Ratio \\
FFT         & Fast Fourier Transform \\
RELAX       & RELAXation method \\
\hline
\end{tabular}
\caption{Acronyms used in this paper}\label{tab:acronyms}
\end{table}

\section{Problem Formulation}\label{sec:PF}
\subsection{One-Bit UWB Radar System}
\begin{figure}[htbp]
\centering
\subfloat[]{\includegraphics[width=0.5\textwidth]{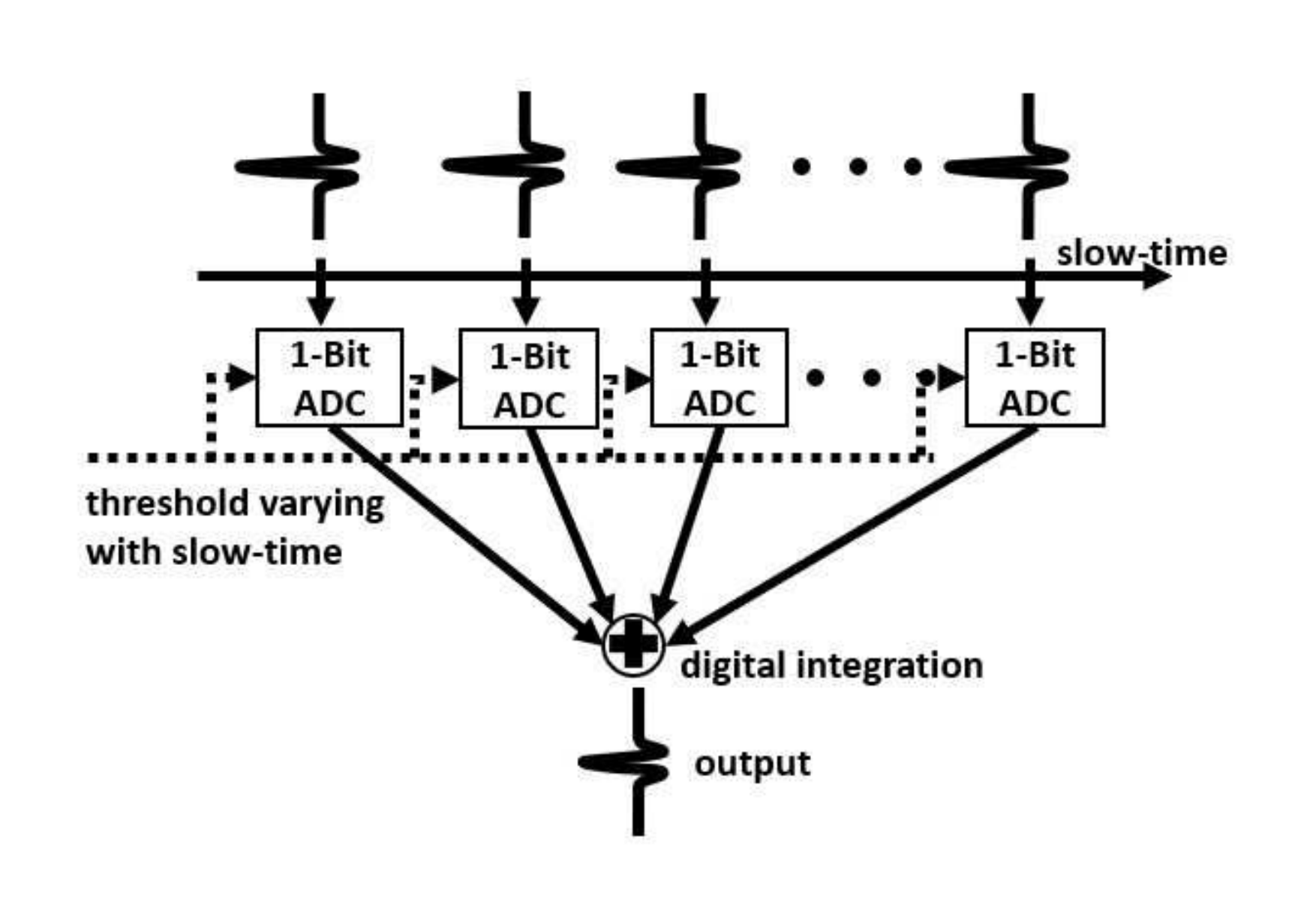}}\\
\subfloat[]{\includegraphics[width=0.5\textwidth]{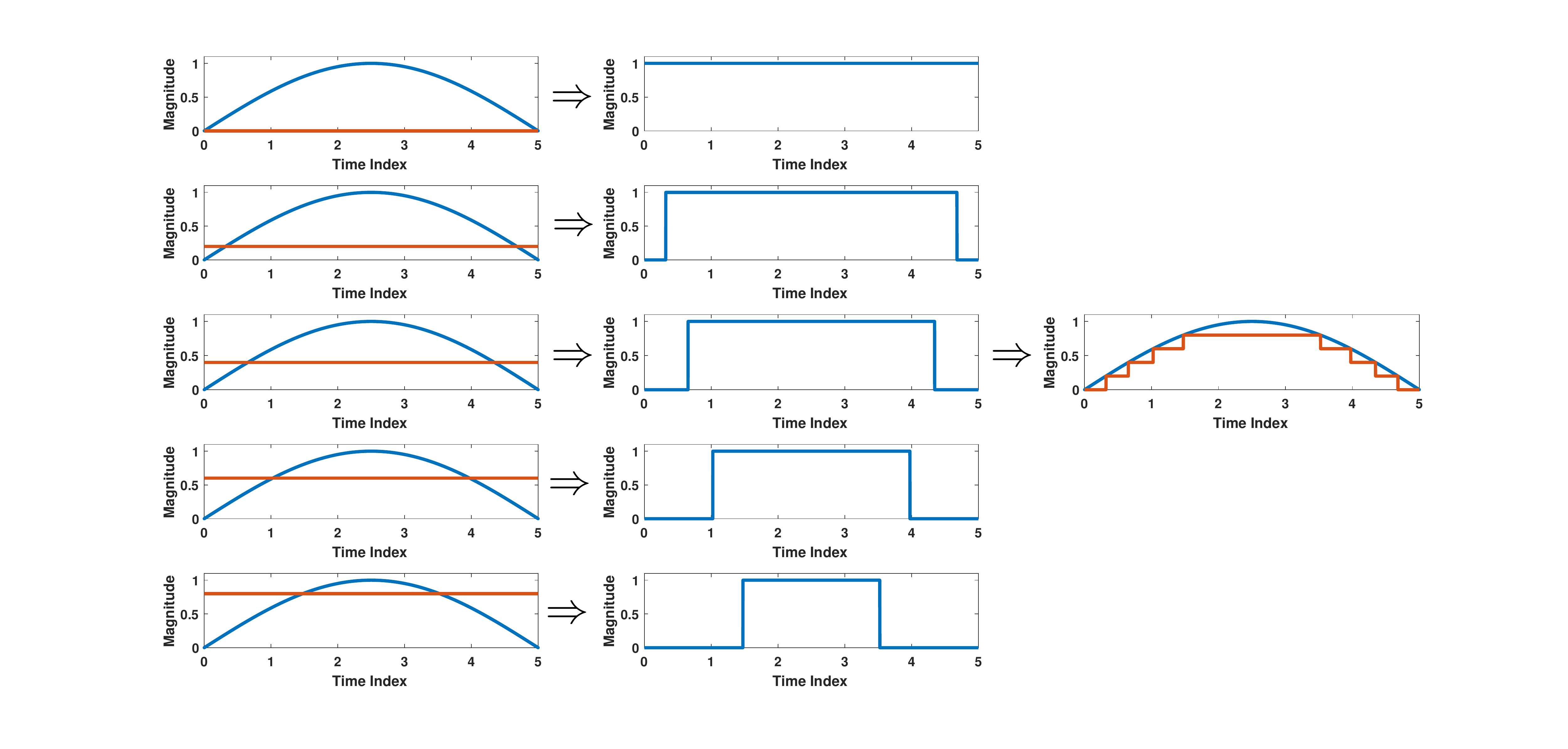}}
\caption{a) The structure of the receiver of a one-bit UWB radar system. Note that in fact there is only one ADC in a one-bit UWB radar sampling different PRIs. b) illustration of the DI method.}
\label{fig:structure}
\end{figure}

We consider a recently-developed impulse UWB radar on a single chip, namely the NVA6100 system \cite{Xethru}. By using the CTBV sampling technique, it can achieve an ultra-high sampling rate of 39 GHz and a 13-bit quantization precision with rather simple hardware \cite{Xethru}, making it a low cost and low power system. The radar transmits a super narrow pulse repeatedly, and at the receiver uses one-bit ADCs to obtain signed measurements. More specifically, each one-bit ADC compares the received signal with a known threshold varying with slow-time. The threshold varies linearly over different PRIs within the CPI, and each one-bit ADC records whether the data samples are larger or smaller than the threshold. Thus, in the absence of RFI and other disturbances, the signed measurement matrix obtained by NVA6100 can be expressed as follows:
\begin{equation}\label{equ:1bnoRFImodel}
\begin{split}
{\bm Z} &= {\rm sign}({\bm S}-{\bm H})\in\mathbb{R}^{N\times M},
\end{split}\end{equation}
where $N$ and $M$ denote the number of fast-time samples per PRI and the number of PRIs or slow-time samples within the CPI, respectively, ${\bm S}$ denotes the radar echo signal and ${\bm H}$ denotes the known threshold matrix. Since the threshold varies linearly with slow-time, each column of the matrix ${\bm H}$ can be expressed as ${\bm h}_m = [-h_{\max}+2(m-1)h_{\max}/(M-1)]{\bm 1}_N, m=1, \dots, M$. In (\ref{equ:1bnoRFImodel}), ${\rm sign}(\cdot)$ is the element-wise sign operator defined as:
\begin{equation}\begin{split}
{\rm sign}(x) = \begin{cases}1, &x\geq 0, \cr -1, &x<0. \end{cases}
\end{split}\end{equation}
Due to the ultra-high sampling rate, the time interval of the CTBV sampling process is also extremely short and it can be assumed that the slow-time samples of the radar echo will be nearly the same in such a short time interval, i.e., ${\bm s}_{:,1} = {\bm s}_{:,2}  = \dots =  {\bm s}_{:,M}  = {\bm s}$. Then the simple digital integration (DI) method \cite{HWLLM07} can be used to recover the radar echo from the signed measurement matrix ${\bm Z}$. The output of the one-bit system using the DI method, $\hat{\bm s}^{\rm DI}$, can be written in the following form:
\begin{align}
\hat{s}^{\rm DI}_n = \left[\Delta h \sum_{m=1}^M \frac{1}{2}\left(z_{n,m}+1\right)\right]-h_{\max}-\Delta h, \\
\Delta h = 2h_{\max}/(M-1), n = 1,\dots,N. \nonumber
\end{align}
The structure of the NVA6100 receiver and an illustration of the DI method are shown in Figure \ref{fig:structure}.
More details about NVA6100, the CTBV sampling technique and the DI method can be found in \cite{HWLLM07, Xethru}. The DI method cannot provide a satisfactory performance when the RFI is strong since DI treats RFI as noise and does not exploit its structure of RFI. Since strong RFI sources exist in practical applications, an effective RFI mitigation technique is needed for the proper operation of one-bit UWB radar systems. An existing RFI mitigation technique for the one-bit UWB radar system, introduced in \cite{ZRGL19,ZRLL20}, is to first use the 1bMMRELAX-1bBIC algorithm \cite{RZLS19} to estimate the RFI parameters and then employ a sparse method for echo recovery. However, this method suffers from high computational complexity since the 1bMMRELAX algorithm that estimates the parameters of the RFI sources requires many iterations. Also, this technique performs the RFI parameter estimation and radar echo recovery separately, which can limit its effectiveness for RFI mitigation and accurate echo recovery.
\subsection{Data Model}
For the NVA6100 one-bit UWB radar system, the received data matrix in the presence of RFI and other disturbances can be written as follows:
\begin{align}\label{equ:1bmodel}
{\bm Z} &= {\rm sign}({\bm Y}-{\bm H})\nonumber \\
        &= {\rm sign}({\bm F}+{\bm S}+{\bm E}-{\bm H})\in\mathbb{R}^{N\times M},
\end{align}
where ${\bm Y}$ is the high-precision data matrix, ${\bm F}$ denotes the RFI matrix and ${\bm E}$ denotes noise and other disturbances. In (\ref{equ:1bmodel}), ${\bm S}$ denotes the radar echo signal, which as already stated is assumed to be invariant over the PRIs, i.e., ${\bm s}_{:,1} = {\bm s}_{:,2}  = \dots =  {\bm s}_{:,M}  = {\bm s}$. Our goal is to recover the radar echo ${\bm s}$ from the signed measurement matrix ${\bm Z}$ while mitigating the impact of the strong RFI and other disturbances.

\section{Review of Weighted SPICE for \\ High-precision Data}\label{sec:c_spice}
In this section, we give a brief review of the original weighted SPICE framework \cite{SZL2014} for high-precision measurements. This review will be beneficial to later interpretation and analysis of the one-bit weighted SPICE counterpart.
Consider sparse parameter estimation for the following linear model:
\begin{equation}
{\bm y} = {\bm B}{\bm x}+{\bm e}\in\mathbb{C}^{N}, \ {\bm B}\in\mathbb{C}^{N\times K},
\end{equation}
where ${\bm x}\in\mathbb{C}^K$ denotes the unknown sparse parameter vector ($K>N$), ${\bm B}$ is the given matrix of regressors and ${\bm e}$ is the zero-mean noise. The covariance matrix of the data vector ${\bm y}$ can be written as
\begin{align}
{\bm R} &= E\{{\bm y}{\bm y}^H\} = E\{({\bm B}{\bm x}+{\bm e})({\bm x}^H{\bm B}^H+{\bm e}^H)\}\nonumber \\
    &={\bm B}E\{{\bm x}{\bm x}^H\}{\bm B}^H+E\{{\bm e}{\bm e}^H\}={\bm B}{\bm P}_x{\bm B}^H+{\bm P}_e \nonumber \\
    & = {\bm A}{\bm P}_{xe}{\bm A}^H,
\end{align}
where
\begin{equation}
{\bm A}=[{\bm B}\ {\bm I}_N], {\bm P}_{xe}={\rm diag}({\bm p}_{xe}), {\bm p}_{xe} = [{\bm p}_x^T, {\bm p}_e^T]^T,
\end{equation}
and where ${\bm P}_x={\rm diag}({\bm p}_x)$, with ${\bm p}_x = [p_1, p_2, \dots, p_K]^T$,  ${\bm P}_e={\rm diag}({\bm p}_e)$, with ${\bm p}_e = [p_{K+1}, p_{K+2}, \dots, p_{K+N}]^T$.
The SPICE algorithm is a covariance fitting approach with the following objective function \cite{SZL2014}:
\begin{equation}
\min_{{\bm p}_{xe}} \|{\bm R}^{-1/2}({\bm y}{\bm y}^H-{\bm R})\|_2^2.
\end{equation}
After some simple calculations, the objective function can be reformulated as \cite{SZL2014}:
\begin{align}\label{equ:c_spice2}
\min_{{\bm p}_{xe}} f_{\rm SPICE}({\bm p}_{xe}) &= {\bm y}^H{\bm R}^{-1}{\bm y}+{\rm tr}\{{\bm R}\}\nonumber \\
&={\bm y}^H{\bm R}^{-1}{\bm y}+\sum_{k=1}^{N+K}w_kp_k
\end{align}
where ${\rm tr}(\cdot)$ denotes the trace of a matrix and $w_k = \|{\bm a}_{:,k}\|_2^2$. The convex optimization problem (\ref{equ:c_spice2}) can be efficiently solved by iteratively minimizing a majorizing function\cite{SZL2014}. At the $(t+1)$-th iteration, the majorizing function of $f_{\rm SPICE}({\bm p}_{xe})$ that will be used has the following form (see \cite{SZL2014,SBL11,SBL11-2,SB12} for details):
\begin{equation}\label{equ:c_spice_mm}
f_{\rm SPICE}({\bm p}_{xe})\leq (\breve{\bm x}^{(t+1)})^H{\bm P}_{xe}^{-1}\breve{\bm x}^{(t+1)}+\sum_{k=1}^{N+K}w_kp_k,
\end{equation}
with
\begin{align}
\breve{\bm x}^{(t+1)} &= \hat{\bm P}^t_{xe}{\bm A}^H(\hat{\bm R}^t)^{-1}{\bm y} \nonumber \\
            &= \begin{bmatrix}\hat{\bm P}^t_{x}{\bm B}^H(\hat{\bm R}^t)^{-1}{\bm y}\\ \hat{\bm P}^t_{e}(\hat{\bm R}^t)^{-1}{\bm y}\end{bmatrix},
\end{align}
where $\hat{\bm P}^t_{xe}$ and $\hat{\bm R}^t = {\bm A}\hat{\bm P}^t_{xe}{\bm A}^H$ denote the estimates of ${\bm P}_{xe}$ and ${\bm R}$ at the $t$-th iteration. Note that $\hat{\bm x}^{(t+1)}=\hat{\bm P}^t_{x}{\bm B}^H(\hat{\bm R}^t)^{-1}{\bm y}$ is the linear minimum mean-squared error (LMMSE) estimate of ${\bm x}$ corresponding to $\hat{\bm p}^t_{x}$ \cite{SZL2014}. The above majorizing function can be minimized easily and ${\bm p}_{xe}$ can be updated by using the following closed-form expression at the $(t+1)$-th iteration:
\begin{align}
\hat{p}_k^{(t+1)} &= |\breve{x}_k^{(t+1)}|/\sqrt{w_k} \nonumber \\
                &=\hat{p}_k^{t}|{\bm a}_{:,k}^H(\hat{\bm R}^t)^{-1}{\bm y}|/\sqrt{w_k}, \quad k=1,\dots,K+N.
\end{align}
By changing the weight vector $\{w_k\}_{k=1}^{K+N}$ in $f_{\rm SPICE}({\bm p}_{xe})$, different weighted SPICE algorithms can be derived, including LIKES and IAA.

The LIKES criterion has the following form \cite{SZL2014}:
\begin{equation}
f_{\rm LIKES}({\bm p}_{xe}) = {\bm y}^H{\bm R}^{-1}{\bm y}^H+\ln|{\bm R}|,
\end{equation}
and its majorizing function at $(t+1)$-th iteration can be written as:
\begin{equation}\label{equ:c_likes_mm}
f_{\rm LIKES}({\bm p}_{xe})\leq (\breve{\bm x}^{(t+1)})^H{\bm P}_{xe}^{-1}\breve{\bm x}^{(t+1)}+\sum_{k=1}^{N+K}w_kp_k,
\end{equation}
with $w_k = {\bm a}_{:,k}^H(\hat{\bm R}^t)^{-1}{\bm a}_{:,k}$. Note that (\ref{equ:c_likes_mm}) and (\ref{equ:c_spice_mm}) have a similar form but with different weights. The IAA algorithm can also be interpreted as a form of weighted SPICE by only setting the weights $w_k$ as ${\hat p}_k^t({\bm a}_{:,k}^H(\hat{\bm R}^t)^{-1}{\bm a}_{:,k})^2$ at the $(t+1)$-th iteration \cite{SZL2014} (for more details about the weighted SPICE framework, we refer to \cite{SZL2014, SLL20-2}).

\section{One-Bit Weighted SPICE for One-bit UWB Radar Systems}\label{sec:1bSPICE}
Considering the sparsity of the RFI sources in the fast-frequency domain, as illustrated in Figure \ref{fig:spec_RFI_mea}, we model the RFI as a sum of sinusoids. Thus we can write the RFI matrix in the following structured form:
\begin{align}
&{\bm F} = {\bm A}_1{\bm X}_1,\quad {\bm A_1}\in\mathbb{C}^{N\times K_1}, \\
&({\bm a}_1)_{:,k} = [1,e^{j\omega_{k}},\dots,e^{j(N-1)\omega_{k}}]^T,\quad k=1,\dots,K_1. \nonumber
\end{align}
Here ${\bm A}_1=[({\bm a}_1)_{:,1},\dots, ({\bm a}_1)_{:,K_1}]$ is the Fourier matrix corresponding to the normalized frequencies $\{\omega_k\}_{k=1}^{K_1}$, which form a grid of $K_1$ points covering the interval $[-\pi,\pi]$. We assume that the grid is fine enough such that the frequencies (normalized by the sampling frequency) corresponding to the RFI sources are on or close enough to the grid points. The elements of the matrix ${\bm X}_1\in\mathbb{C}^{K_1\times M}$ denote the complex amplitudes of the corresponding frequency points in the grid; this matrix has a row sparse property due to the sparsity of the RFI sources in the fast-frequency domain. Note that since ${\bm R}$ is a real-valued matrix, $(x_1)_{K_1-k+1,m}$ should be the complex conjugate of $(x_1)_{k,m}, m=1,\dots,M$.

The radar echo matrix ${\bm S}$ can also be rewritten as:
%\begin{equation}\begin{split}
%&{\bm S} = {\bm A}_2{\bm X}_2,\ ({\bm x}_2)_1 = \dots = ({\bm x}_2)_M = {\bm x}_2\in\mathbb{R}^{K_2\times 1},\\
%&{\bm A}_2 = \begin{bmatrix}
%\psi(0)&\psi(-\frac{N}{K_2}\Delta t)& \cdots&\psi(-N\Delta t)\\
%\psi(\Delta t) &\psi((1-\frac{N}{K_2})\Delta t)&\cdots&\psi((1-N\Delta) t)\\
%\vdots &\vdots&\ddots&\vdots\\
%\psi(N\Delta t) &\psi((N-\frac{N}{K_2})\Delta t)&\cdots&\psi(0)
%\end{bmatrix}\in\mathbb{R}^{N\times K_2},
%\end{split}
%\end{equation}
\begin{equation}
{\bm S} = {\bm A}_2{\bm X}_2,\ ({\bm x}_2)_{:,1} = \dots = ({\bm x}_2)_{:,M} = {\bm x}_2\in\mathbb{R}^{K_2\times 1},\\
\end{equation}
where ${\bm A}_2$ denotes the dictionary whose columns are time-shifted, digitized versions of the transmitted impulse $\psi(t)$. Each column of the dictionary ${\bm A}_2$ can be thought of as the discrete version of the time-shifted analog signal $\psi
\left(t-\frac{N\Delta t}{K_2}k\right),\ k=1,2,\dots, K_2$, with $\Delta t$ denoting the sampling interval. ${\bm x}_2$ is the vector containing the information on the magnitudes and positions of the radar echoes. Due to the sparsity of the radar echoes in the fast-time domain, the vector ${\bm x}_2$ also possesses a sparse property. Thus, we can extend the idea of the weighted SPICE framework, which exploits the sparse property of ${\bm X}_1$ and ${\bm X}_2$, to solve the joint RFI mitigation and echo recovery problem for the one-bit UWB radar system. Inspired by \cite{SZL2014}, a one-bit weighted SPICE based framework was presented in \cite{SLL20-2} for parameter estimation in automotive radar systems. However, the method in \cite{SLL20-2} cannot be directly applied to our one-bit RFI mitigation and echo recovery problem. In this paper, we extend the work in \cite{SLL20-2} and develop a new one-bit weighted SPICE methodology for joint RFI mitigation and radar echo recovery for the one-bit UWB radar system.

To use the weighted SPICE framework for signed measurements, a penalty term for the sign disagreement between the estimated signal and the signed measurements is necessary. Here we consider using the negative log-likelihood function of the signal model due to the good performance of the maximum likelihood estimator. We assume that the elements of the noise, i.e., ${\bm E}$, obeys an i.i.d. Gaussian distribution with zero-mean and unknown variance $\sigma^2$. The numerical and experimental examples in Section \ref{sec:experiment} will show that the proposed algorithms are robust to this assumption. Then the negative log-likelihood function of ${\bm Z}$ in (\ref{equ:1bmodel}) can be written as:
\begin{align}
&L({\bm X}_1, {\bm x}_2, \sigma)\nonumber \\
&=\!\!-\!\!\sum_{m=1}^M\!\!\sum_{n=1}^N\ln\Phi\left(\!\!z_{n,m}\frac{({\bm a}_1)_{n,:}({\bm x}_1)_{:,m}+({\bm a}_2)_{n,:}{\bm x}_2-h_{n,m}}{\sigma}\!\!\right) \nonumber \\
&=\!\!-\!\!\!\sum_{m=1}^M\!\!\sum_{n=1}^N\ln\Phi\!\left(z_{n,m}\left[({\bm a}_1)_{n,:}(\tilde{\bm x}_1)_{:,m}\!+\!({\bm a}_2)_{n,:}\tilde{\bm x}_2-\eta h_{n,m}\right]\right),
\end{align}
where $\Phi(x)$ denotes the cumulative distribution function of the standard normal distribution, and $\tilde{\bm X}_1 = {\bm X}_1/\sigma, \tilde{\bm x}_2 = {\bm x}_2/\sigma, \eta = 1/\sigma$. Then, similar to Section \ref{sec:c_spice}, considering the sparse properties of the signal model we introduce the following objective function for the joint one-bit RFI mitigation and echo recovery:
\begin{align}\label{equ:1bspice}
\min_{\tilde{\bm X}_1, \tilde{\bm X}_2,\eta, {\bm p}_1, {\bm p}_2}\!\!&\!\!\!-\!\!\!\sum_{m=1}^M\!\sum_{n=1}^N\!\ln\Phi(\!z_{n,m}[({\bm a}_1)_{n,:}(\tilde{\bm x}_1)_{:,m}\!\!+\!\!({\bm a}_2)_{n,:}(\tilde{\bm x}_2)_{:,m} \nonumber \\
&-\eta h_{n,m}])\!+\!\!\!\sum_{m=1}^M (\tilde{\bm x}_1)_{:,m}^H{\bm P}_1^{-1}(\tilde{\bm x}_1)_{:,m}\nonumber \\
&+\!\!\!\sum_{m=1}^M (\tilde{\bm x}_2)_{:,m}^H{\bm P}_2^{-1}(\tilde{\bm x}_2)_{:,m}+\xi{\bm w}_1^T {\bm p_1}+ {\bm w}_2^T {\bm p_2}, \nonumber \\
{\rm s.t.} &\ (\tilde{\bm x}_2)_{:,1} = \dots = (\tilde{\bm x}_2)_{:,M}=\tilde{\bm x}_2,
\end{align}
where ${\bm P}_1 = {\rm diag}({\bm p_1}), {\bm P}_2 = {\rm diag}({\bm p_2})$. In (\ref{equ:1bspice}), ${\bm w}_1$ and ${\bm w}_2$ denote the weight vectors of three SPICE methods discussed in Section \ref{sec:c_spice}, shown in Table \ref{tab:1bweight}, and $\xi$ is a penalty factor used to balance the power of the estimated RFI and radar echoes.

Since the problem in (\ref{equ:1bspice}) is difficult to solve directly, we use the majorization-minimization (MM) \cite{HL04,SS04} technique to simplify it. By using the MM technique, the objective function in (\ref{equ:1bspice}) can be majorized as in (\ref{equ:1bspice_mm}) at the $(t+1)$-th MM iteration (see \cite{RZLS19}):

\begin{align}\label{equ:1bspice_mm}
\min_{\tilde{\bm X}_1, \tilde{\bm X}_2,\eta, {\bm p}_1, {\bm p}_2}&\frac{1}{2}\!\!\sum_{m=1}^M\!\!\left \|\!\left({\bm A}_1(\tilde{\bm x}_1)_{:,m}\!\!+\!\!{\bm A}_2(\tilde{\bm x}_2)_{:,m}\!\!-\!\!(\eta{\bm h}_{:,m}\!\!+\!\!{\bm g}_{:,m}^t)\!\right)\!\right \|_2^2 \nonumber \\
&\!\!+\!\!\!\!\sum_{m=1}^M\!(\tilde{\bm x}_1)_{:,m}^H{\bm P}_{1}^{-1}
\!(\tilde{\bm x}_1)_{:,m}\!\!+\!\!\!\!\sum_{m=1}^M\!(\tilde{\bm x}_2)_{:,m}^H{\bm P}_{2}^{-1}\!(\tilde{\bm x}_2)_{:,m} \nonumber\\
&\!\!+\xi{\bm w}_{1}^T {\bm p}_{1}+ {\bm w}_2^T {\bm p}_{2}, \nonumber \\
{\rm s.t.}&\ (\tilde{\bm x}_2)_{:,1} = \dots = (\tilde{\bm x}_2)_{:,M}=\tilde{\bm x}_2,
\end{align}
where
\begin{align}
\label{equ:mm_aux}
&{\bm g}_{:,m}^t = {\bm z}_{:,m}\odot({\bm \gamma}_{:,m}^t-f'({\bm \gamma}_{:,m}^t)),\\
&f'(x) = -\frac{\phi(x)}{\Phi(x)}, \\
&{\bm \gamma}_{:,m}^t = {\bm z}_{:,m}\odot\left({\rm Re}\left[{\bm A}_1(\hat{\tilde{\bm x}}_1^t)_{:,m}+{\bm A}_2\hat{\tilde{\bm x}}_2^t\right] - \hat{\eta}^t{\bm h}_{:,m} \right).
\end{align}
Here, $\hat{\tilde{\bm X}}_1^t$, $\hat{\tilde{\bm x}}_2^t$ and $\hat{\eta}^t$ denote the estimates of the corresponding variables at the $t$-th MM iteration. ${\rm Re}(\cdot)$ means taking the real part of the corresponding variable and $\phi(x)$ denotes the probability density function of the standard normal distribution.
To simplify the calculation, we rewrite (\ref{equ:1bspice_mm}) in the following form:
\begin{align}\label{equ:1bspice_mm2}
\min_{\tilde{\bm X}_1, \tilde{\bm X}_2,\eta, {\bm p}_1, {\bm p}_2} &\frac{1}{2}\sum_{m=1}^M \left \|\left({\bm A}\tilde{\bm x}_{:,m} - (\eta{\bm h}_{:,m}+{\bm g}_{:,m}^t)\right) \right \|_2^2 \nonumber \\
&+\!\!\sum_{m=1}^M \tilde{\bm x}_{:,m}^H{\bm P}^{-1}\tilde{\bm x}_{:,m}+ \xi{\bm w}_1^T {\bm p}_{1}+ {\bm w}_2^T {\bm p}_{2}, \nonumber \\
{\rm s.t.} &\ (\tilde{\bm x}_2)_{:,1} = \dots = (\tilde{\bm x}_2)_{:,M}=\tilde{\bm x}_2,
\end{align}
where
\begin{equation}
{\bm A}\!=\![{\bm A}_1, {\bm A}_2], \tilde{\bm x}_{:,m}\!=\!\begin{bmatrix}(\tilde{\bm x}_1)_{:,m} \\ (\tilde{\bm x}_2)_{:,m}\end{bmatrix}, {\bm P}\!=\!{\rm diag}({\bm p}), {\bm p}\!=\!\begin{bmatrix}{\bm p}_1 \\ {\bm p}_2\end{bmatrix}.
\end{equation}

We minimize (\ref{equ:1bspice_mm2}) by cyclicly updating one group of variables while fixing the rest.

\subsection{Update of $\{\tilde{\bm x}_{:,m}\}_{m=1}^M$ and $\eta$}
To update $\{\tilde{\bm x}_{:,m}\}_{m=1}^M$ and $\eta$, we set to zero the derivatives of (\ref{equ:1bspice_mm2}) with respect to $\{\tilde{\bm x}_{:,m}\}_{m=1}^M$ and $\eta$. We obtain the following equation for the $(t+1)$-th MM iteration:
\begin{equation}
{\bm A}^H\left({\bm A}\tilde{\bm x}_{:,m}-(\eta{\bm h}_{:,m}+{\bm g}_{:,m}^t)\right)+2(\hat{\bm P}^t)^{-1}\tilde{\bm x}_{:,m}={\bm 0},
\end{equation}
which gives:
\begin{equation}
\tilde{\bm x}_{:,m} = \left({\bm A}^H{\bm A}+2(\hat{\bm P}^t)^{-1}\right)^{-1}{\bm A}^H\left(\eta{\bm h}_{:,m}+{\bm g}_{:,m}^t\right).
\end{equation}
Using the matrix inversion lemma yields:
\begin{equation}
\tilde{\bm x}_{:,m} =\begin{bmatrix}(\tilde{\bm x}_1)_{:,m}\\ (\tilde{\bm x}_2)_{:,m}\end{bmatrix}=\hat{\bm P}^t{\bm A}^H(\hat{\bm R}^t)^{-1}\left(\eta{\bm h}_{:,m}+{\bm g}_{:,m}^t\right),
\end{equation}
where
\begin{equation}\label{equ:updata_x_1}
\hat{\bm R}^t={\bm A}\hat{\bm P}^t{\bm A}^H+2{\bm I}_N,
\end{equation}
with $\hat{\bm P}^t$ denoting the estimate of ${\bm P}$ from the $t$-th MM iteration.

We also have
\begin{equation}\label{equ:update_eta_1}
\begin{split}
\sum_{m=1}^M {\bm h}_{:,m}^T\left(\eta{\bm h}_{:,m}+{\bm g}_{:,m}^t - {\bm A}\tilde{\bm x}_{:,m}\right)=0.
\end{split}\end{equation}
Substituting the expression of $\tilde{\bm x}_m$ in (\ref{equ:updata_x_1}) into (\ref{equ:update_eta_1}), we obtain:
\begin{equation}
\sum_{m=1}^M\!{\bm h}_{:,m}^T\!\left(\!\eta{\bm h}_{:,m}\!+\!{\bm g}_{:,m}^t\!\!-\!\!{\bm A}\hat{\bm P}^t{\bm A}^H(\hat{\bm R}^t)^{-1}\!\!\left(\eta{\bm h}_{:,m}+{\bm g}_{:,m}^t\right)\!\right)\!=\!0,
\end{equation}
which yields:
\begin{equation}
\hat{\eta}^{(t+1)}=\max\left(0,\frac{\sum_{m=1}^M{\bm h}_{:,m}^T(\hat{\bm R}^t)^{-1}{\bm g}_{:,m}^t}{\sum_{m=1}^M{\bm h}_{:,m}^T(\hat{\bm R}^t)^{-1}{\bm h}_{:,m}}\right).
\end{equation}
Next, we insert the $\hat{\eta}^{(t+1)}$ above into (\ref{equ:updata_x_1}) and obtain the update formula of $\tilde{\bm x}_{:,m}$:
\begin{equation}\label{equ:updata_x_2}
\hat{\tilde{\bm x}}_{:,m}^{(t+1)}\!=\!\!\begin{bmatrix}(\hat{\tilde{\bm x}}_1^{(t+1)})_{:,m}\\ (\hat{\tilde{\bm x}}_2^{(t+1)})_{:,m}\end{bmatrix}\!\!=\!\hat{\bm P}^t{\bm A}^H(\hat{\bm R}^t)^{-1}\!\left(\hat{\eta}^{(t+1)}{\bm h}_{:,m}+{\bm g}_{:,m}^t\right).
\end{equation}
Considering the constraint $(\tilde{\bm x}_2)_{:,1} = \dots = (\tilde{\bm x}_2)_{:,M}=\tilde{\bm x}_2$, we have:
\begin{equation}
\hat{\tilde{\bm x}}_{2}^{(t+1)}=\frac{1}{M}\sum_{m=1}^M (\hat{\tilde{\bm x}}_2^{(t+1)})_{:,m}\ .
\end{equation}

Note that, if we interpret $\hat{\eta}^{(t+1)}{\bm h}_{:,m}+{\bm g}_{:,m}^t$ as high-precision ``input data", the update formula of $\tilde{\bm x}_{:,m}$ is similar to the one we used in the weighted SPICE framework for high-precision data sets (see Section \ref{sec:c_spice}). Thus, the $(t+1)$-th MM iteration of the one-bit weighted SPICE can be viewed as a weighted SPICE algorithm for the high-precision input $\hat{\eta}^{(t+1)}{\bm h}_{:,m}+{\bm g}_{:,m}^t$ contaminated by zero-mean Gaussian noise with covariance matrix $2{\bm I}_N$ \cite{SLL20-2}. Similar to the weights used in Section \ref{sec:c_spice}, the weights of these one-bit weighted SPICE algorithms at the $(t+1)$-th MM iteration are summarized in Table \ref{tab:1bweight}.

\begin{table}[htbp]
\centering
\resizebox{0.45\textwidth}{!}{
\begin{tabular}{|c|l|}
\hline
          & Weight   ${\bm w}_i,\ i = 1$ or $2$      \\
\hline
1bSPICE   &  $(w_i^{t+1})_k=\left\|({\bm a}_{i})_{:,k}\right\|_2^2$  \\
1bLIKES   &  $(w_i^{t+1})_k=({\bm a}_{i})_{:,k}^H(\hat{\bm R}^t)^{-1}({\bm a}_{i})_{:,k}$  \\
1bIAA     &  $(w_i^{t+1})_k=(p_i)_k^t\left(({\bm a}_{i})_{:,k}^H(\hat{\bm R}^t)^{-1}({\bm a}_{i})_{:,k}\right)^2$  \\
\hline
\end{tabular}}
\caption{Weight vectors ${\bm w}_1$ and ${\bm w}_2$ at the $(t+1)$-th MM iteration of three one-bit weighted SPICE algorithms.}\label{tab:1bweight}
\end{table}

\subsection{Update of ${\bm p}_1, {\bm p}_2$}
To update ${\bm p}_1$ in the $(t+1)$-th MM iteration, we need to solve the following subproblem:
\begin{equation}
\min_{{\bm p}_1} \sum_{m=1}^M(\hat{\tilde{\bm x}}_1^{(t+1)})_{:,m}^H{\bm P}_1^{-1}(\hat{\tilde{\bm x}}_1^{(t+1)})_{:,m}+\xi({\bm w}_{1}^{(t+1)})^T {\bm p}_{1},
\end{equation}
which can be rewritten as:
\begin{equation}
\min_{{\bm p}_1}\sum_{k=1}^{K_1}\!\left(\sum_{m=1}^M\left|(\hat{\tilde{x}}_1^{(t+1)})_{k,m}\right|^2/(p_1)_k+\xi (w_1^{(t+1)})_k(p_1)_k\!\right)\!.
\end{equation}
The above objective function achieves its minimum when:
\begin{align}
\sum_{m=1}^M\left|(\hat{\tilde{x}}_1^{(t+1)})_{k,m}\right|^2/(p_1)_k=\xi (w_1^{(t+1)})_k(p_1)_k,\nonumber \\
k=1,\dots,K_1.
\end{align}
Thus, $\hat{\bm p}_1^{(t+1)}$ can be updated by:
\begin{equation}\label{equ:updata_p1}
(\hat{p}_1^{(t+1)})_k = \sqrt{\frac{\sum_{m=1}^M\left|(\hat{\tilde{x}}_1^{(t+1)})_{k,m}\right|^2}{\xi(w_1^{(t+1)})_k}},\quad k=1,\dots,K_1.
\end{equation}
Similarly, we can update $\hat{\bm p}_2^{(t+1)}$ by:
\begin{equation}
(\hat{p}_2^{(t+1)})_k = \sqrt{\frac{\sum_{m=1}^M\left|(\hat{\tilde{x}}_2^{(t+1)})_{k,m}\right|^2}{(w_2^{(t+1)})_k}},\quad k=1,\dots,K_2,
\end{equation}
which yields:
\begin{equation}\label{equ:updata_p2}
(\hat{p}_2^{(t+1)})_k =\sqrt{\frac{M\left|(\hat{\tilde{x}}_2^{(t+1)})_k\right|^2}{(w_2^{(t+1)})_k}}, \quad k=1,\dots,K_2.
\end{equation}

The one-bit weighted SPICE methodolgy for joint RFI mitigation and echo recovery is summarized in Table \ref{tab:1bspice}. The choice of the user-parameter $\xi$ will be discussed in Section \ref{sec:experiment}.
\begin{table*}[htbp]
\centering
\resizebox{0.7\textwidth}{!}{
\begin{tabular}{|l|l|}
\hline
   Step       &     Operation     \\
\hline
\multirow{2}{*}{1. Initialization} &
\multirow{2}{*}{\begin{tabular}[c]{@{}l@{}}
$\begin{cases}(x_1^0)_{k,m}=1+j, k=1,\dots, K_1/2, \cr (x_1^0)_{k,m}=1-j, k=K_1/2+1,\dots, K_1.\end{cases}$\\
$(x_2^0)_{k,m}=1, k=1,\dots,K_2$.\end{tabular}}\\
          &          \\
          &          \\
          &          \\
\hline
2. Computation of ${\bm R}^t$ and  ${\bm g}_{:,m}^t$        &  See (\ref{equ:updata_x_1}) and (\ref{equ:mm_aux})        \\
\hline
3. Update of weights          &   See Table \ref{tab:1bweight}       \\
\hline
4. Update of $\hat{\eta}^{t+1}$          & $\hat{\eta}^{t+1}=  \max\left(0,\frac{\sum_{m=1}^M{\bm h}_{:,m}^T(\hat{\bm R}^t)^{-1}{\bm g}_{:,m}^t}{\sum_{m=1}^M{\bm h}_{:,m}^T(\hat{\bm R}^t)^{-1}{\bm h}_{:,m}}\right)$       \\
\hline
5. Update of $\hat{\tilde{\bm x}}_{:,m}^{(t+1)}$    & See (\ref{equ:updata_x_2}) \\
\hline
6. Update of $\hat{\bm p}_1^{(t+1)}, \hat{\bm p}_2^{(t+1)}$& See (\ref{equ:updata_p1}) and (\ref{equ:updata_p2})\\
\hline
\multicolumn{2}{|c|}{\begin{tabular}[l]{@{}l@{}}Iterate Steps 2$\sim$6 until $\left|{\bm p}^{(t+1)}-{\bm p}^t\right|_2/\left|{\bm p}^t\right|_2<10^{-6}$ \\
or $t$ reaches a prescribed maximum iteration number $T_M$.\end{tabular}}\\
\hline
\multicolumn{2}{|c|}{Result: $\hat{\tilde{\bm X}}_1, \hat{\tilde{\bm x}}_2, \hat{\eta}$.}\\
\hline
\end{tabular}}
\caption{One-Bit Weighted SPICE for Joint RFI Mitigation and Radar Echo Recovery}\label{tab:1bspice}
\end{table*}

Finally, the estimate $\hat{\bm x}_2$ of ${\bm x}_2$, can be obtained as $\hat{\tilde{\bm x}}_2/\hat{\eta}$, and the recovered radar echo signal as $\hat{\bm s} = {\bm A}_2\hat{\bm x}_2$.
\subsection{Computational Complexity Analysis}
We now compare the computational complexity of the one-bit weighted SPICE framework with that of the separate 1bMMRELAX-1bBIC and radar echo recovery technique \cite{ZRLL20}. For simplicity only the computationally dominating steps are considered herein. First, note that the computational complexity of the one-bit weighted SPICE framework, which mainly comes from the matrix inversion in (\ref{equ:updata_x_2}), is $\mathcal{O}(N^3)$. Consider next the technique in \cite{ZRLL20}. Suppose that $N_1$-point ($N_1\gg N$) zero-padded FFT operations are used in the 1bMMRELAX iterations \cite{ZRLL20} and the number of RFI sources is $Q$; then the computational complexity of 1bMMRELAX is $\mathcal{O}(Q^2MN_1\log N_1)$. For the fast frequency initialization, if MM and Alternating Direction Method of Multipliers (ADMM) \cite{BPCPE11,ZRLL20} approaches are used, the computational complexity is $\mathcal{O}(K_1^3)$ with $K_1\gg N$, which comes from matrix inversion operations. Similarly, the computational complexity of the sparse radar echo recovery is $\mathcal{O}(K_2^3)$ with $K_2\gg N$. Thus the total computational complexity of the separate 1bMMRELAX-1bBIC and sparse radar echo recovery technique is $\mathcal{O}(Q^2MN_1\log N_1+K_1^3+K_2^3)$ which is much higher than that of the joint one-bit weighted SPICE framework.
\section{Simulated and Experimental Examples}\label{sec:experiment}
In this section, we compare the RFI mitigation and echo recovery performance of the joint one-bit weighted SPICE framework with that of the separate 1bMMRELAX-1bBIC and sparse echo recovery technique \cite{ZRGL19,ZRLL20}, and the DI method \cite{HWLLM07}, for the one-bit NVA6100 UWB radar. For simplicity, we will refer to 1bMMRELAX-1bBIC as RELAX hereafter. The computational costs of the one-bit weighted SPICE algorithms and RELAX will also be compared. Depending on the weight vectors used, the one-bit weighted SPICE framework includes 1bLIKES, 1bSPICE and 1bIAA, as summarized in Table \ref{tab:1bweight}.

We conduct experiments using simulated RFI-free UWB radar data and two different RFI data sets: a simulated RFI data set and a measured RFI data set. The measured RFI data set was collected by an ARL radar receiver with its antenna pointing toward Washington DC. More details about the experimental data collection can be found in \cite{NTD14, NT16, RNKWS07}. Because the sampling rate of the ARL radar receiver is $8$ GHz, we assume that all data sets used in this section are obtained at an $8$ GHz sampling rate.

\begin{figure*}[htbp]
\centering
\subfloat[]{\label{fig:pules}\includegraphics[width=0.33\textwidth]{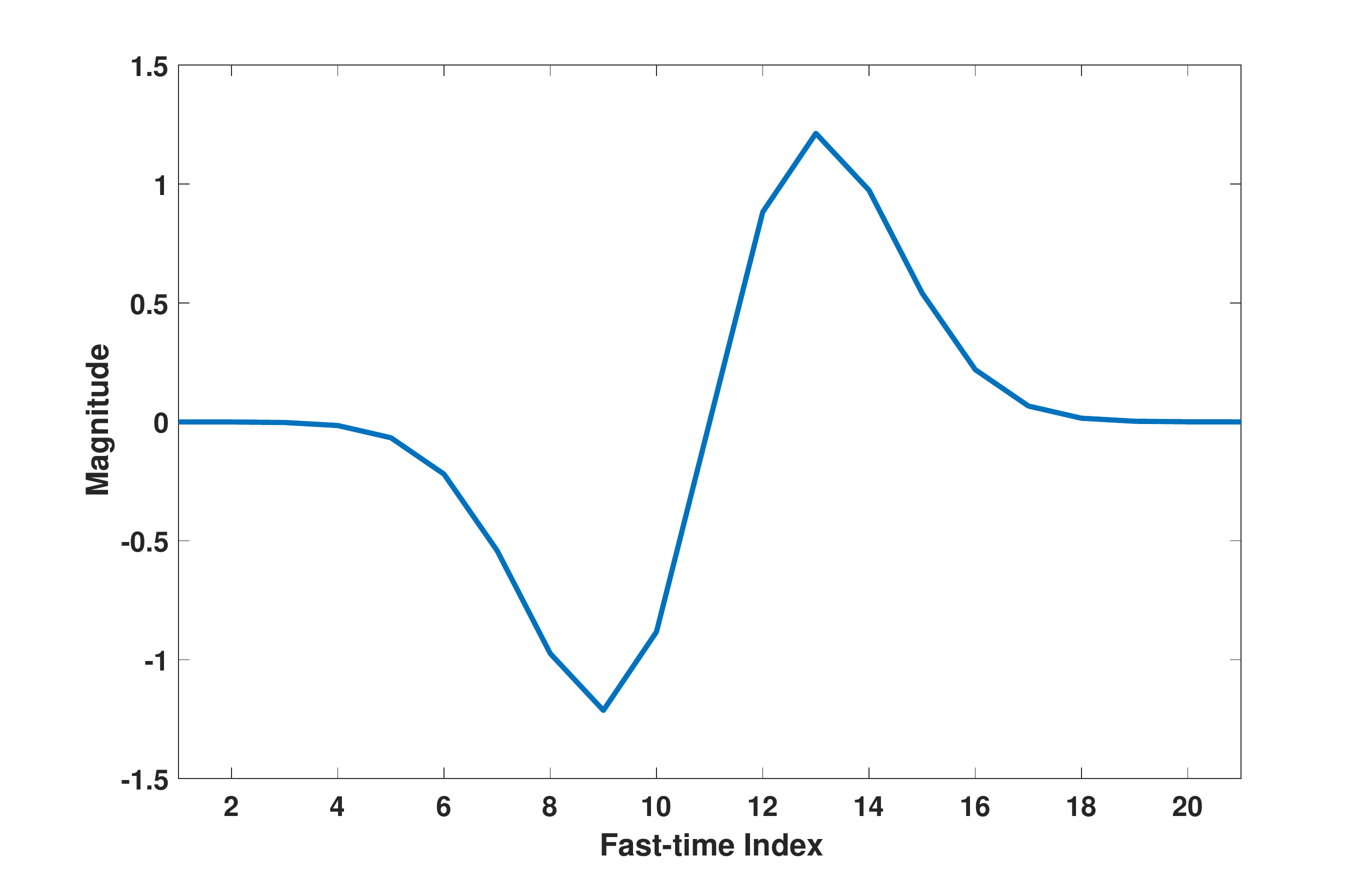}}
\subfloat[]{\label{fig:pules_fft}\includegraphics[width=0.33\textwidth]{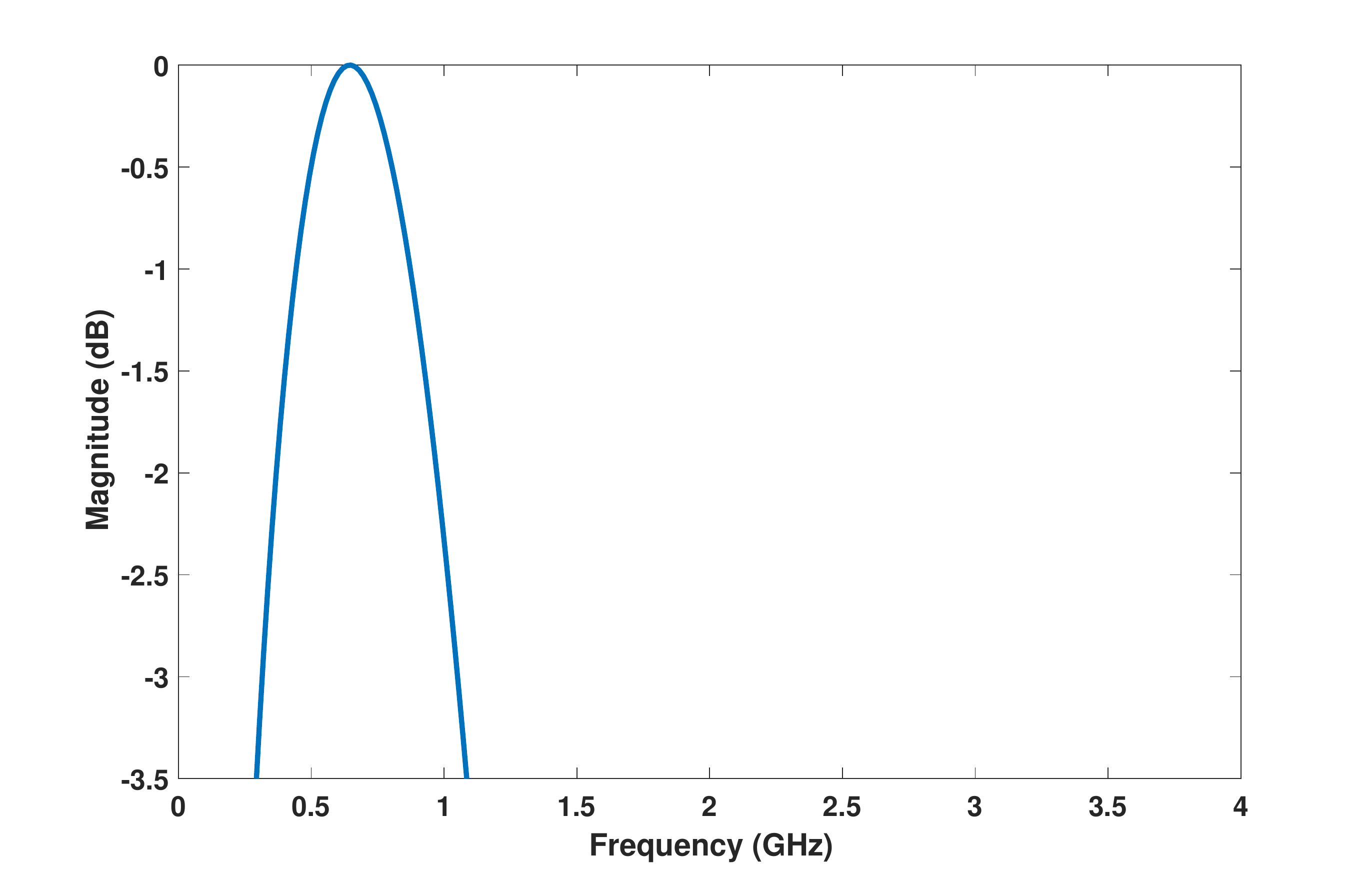}}
\subfloat[]{\label{fig:echo}\includegraphics[width=0.33\textwidth]{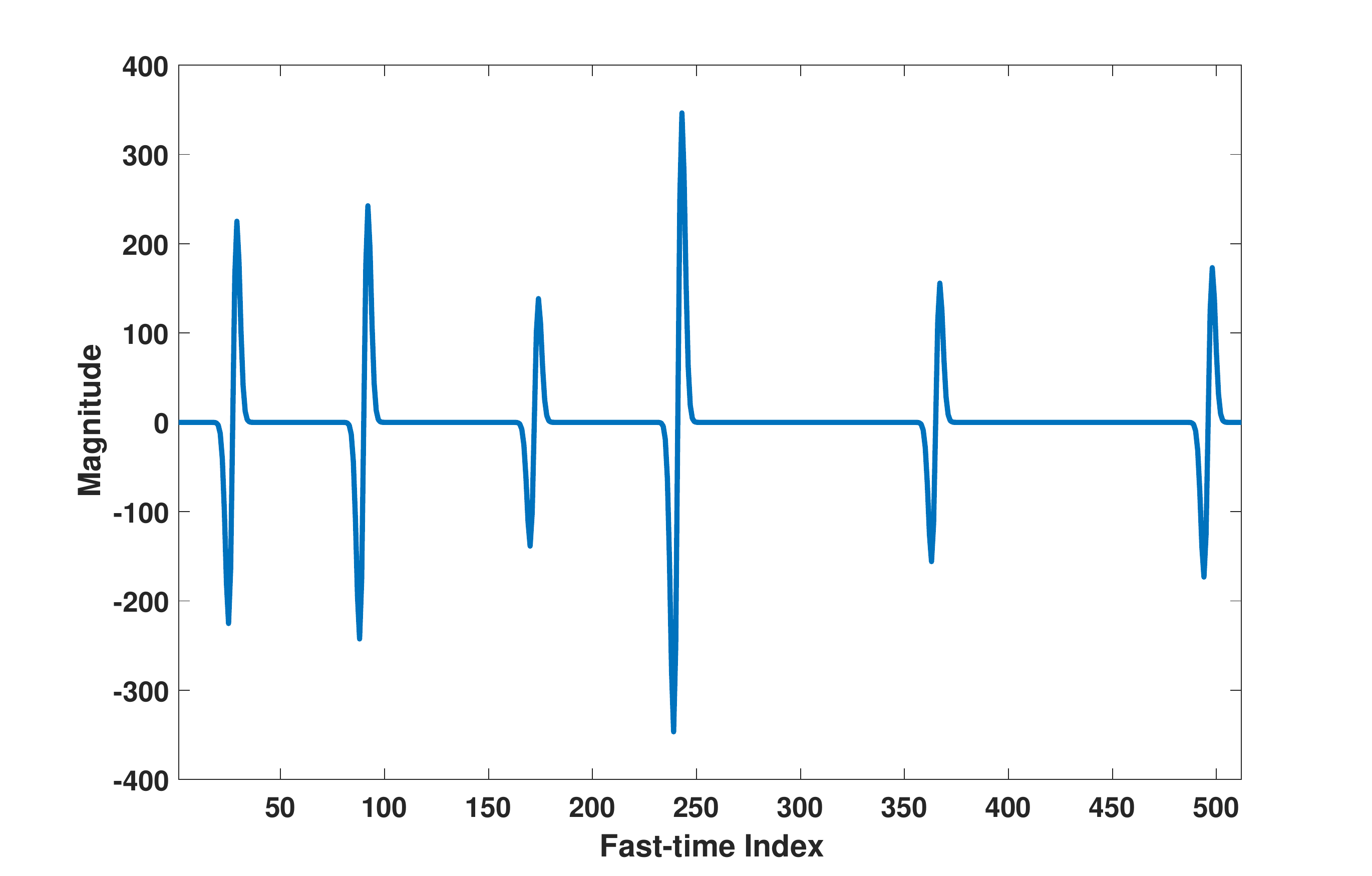}}
\caption{(a) Simulated transmitted radar pulse; (b) Spectrum of the transmitted radar pulse; (c) Simulated RFI-free and noise-free radar echoes for one PRI.}\label{fig:exp_orig}
\end{figure*}

All data sets contain $8192$ slow-time samples within a CPI and $512$ fast-time samples per PRI, i.e., $M=8192,N=512$. The transmitted radar pulse is shaped as the first-order derivative of a Gaussian pulse covering the frequency range of $300\sim 1100$ MHz (see Figures \ref{fig:exp_orig}(a) and \ref{fig:exp_orig}(b)). The simulated radar echoes are generated by using $6$ targets at different ranges with different amplitudes (see Figure \ref{fig:exp_orig}(c)).

All signed measurements are obtained by sampling the RFI-contaminated data via the CTBV sampling technology. All experiments are run using Matlab 2017a installed on a PC with 2.40 GHz CPU and 16 G RAM.

\subsection{Implementation Details}
For the three one-bit weighted SPICE methods, we set $K_1=K_2=4N$ and $\xi = 0.4M$. The maximum iteration number $T_M$ is set to $100$. The implementation details for the RELAX method can be found in \cite{ZRLL20}.

\subsection{Evaluation Metric}
Note that the measured RFI data set inevitably contains noise and other disturbances whereas the simulated RFI data set is noise-free. We add white Gaussian noise ${\bm E}$ to the simulated RFI data set, but we do not add extra noise to the measured (and hence already noisy) RFI data set. Different interference-to-noise ratios (INRs) defined as $20\log_{10}\frac{\|{\bm R}\|_2}{\|{\bm E}\|_2}$ (dB), will be considered in Section \ref{sec:sim_RFI}.

Define the signal-to-interference-plus-noise ratio (SINR) for the simulated RFI data sets as follows:
\begin{equation}
{\rm SINR} = 20\log_{10}\frac{||{\bm S}||_2}{||{\bm R}+{\bm E}||_2}\quad({\rm dB}).
\end{equation}
For the case of measured RFI data sets, which already contain noise, the SINR is computed as:
\begin{equation}
{\rm SINR} = 20\log_{10}\frac{||{\bm S}||_2}{||{\bm R}||_2}\quad({\rm dB}).
\end{equation}
We fix the radar echo signal and add a scaled simulated RFI plus noise or a scaled measured RFI to obtain contaminated data sets with various SINR values. Due to the
low transmit power of the NVA6100 one-bit UWB radar, its signed measurements are commonly contaminated by relatively strong RFI. Therefore, we only consider the severe RFI cases with SINR $\leq -25$ dB in this section.

The maximum threshold $h_{\max}$ is set to $400$ for all cases according to the magnitude of the echo signal in Figure \ref{fig:exp_orig}(c). The echo recovery performance is measured by using the normalized recovery error (NRE):
\begin{equation}
{\rm NRE} = 20\log_{10}\frac{||{\bm s}-\hat{\bm s}||_2}{||{\bm s}||_2}\quad({\rm dB}),
\end{equation}
where $\hat{\bm s}$ is the recovered UWB radar echo signal.
\subsection{Simulated RFI case}\label{sec:sim_RFI}
\begin{figure}[htbp]
\centering
\includegraphics[width=0.45\textwidth]{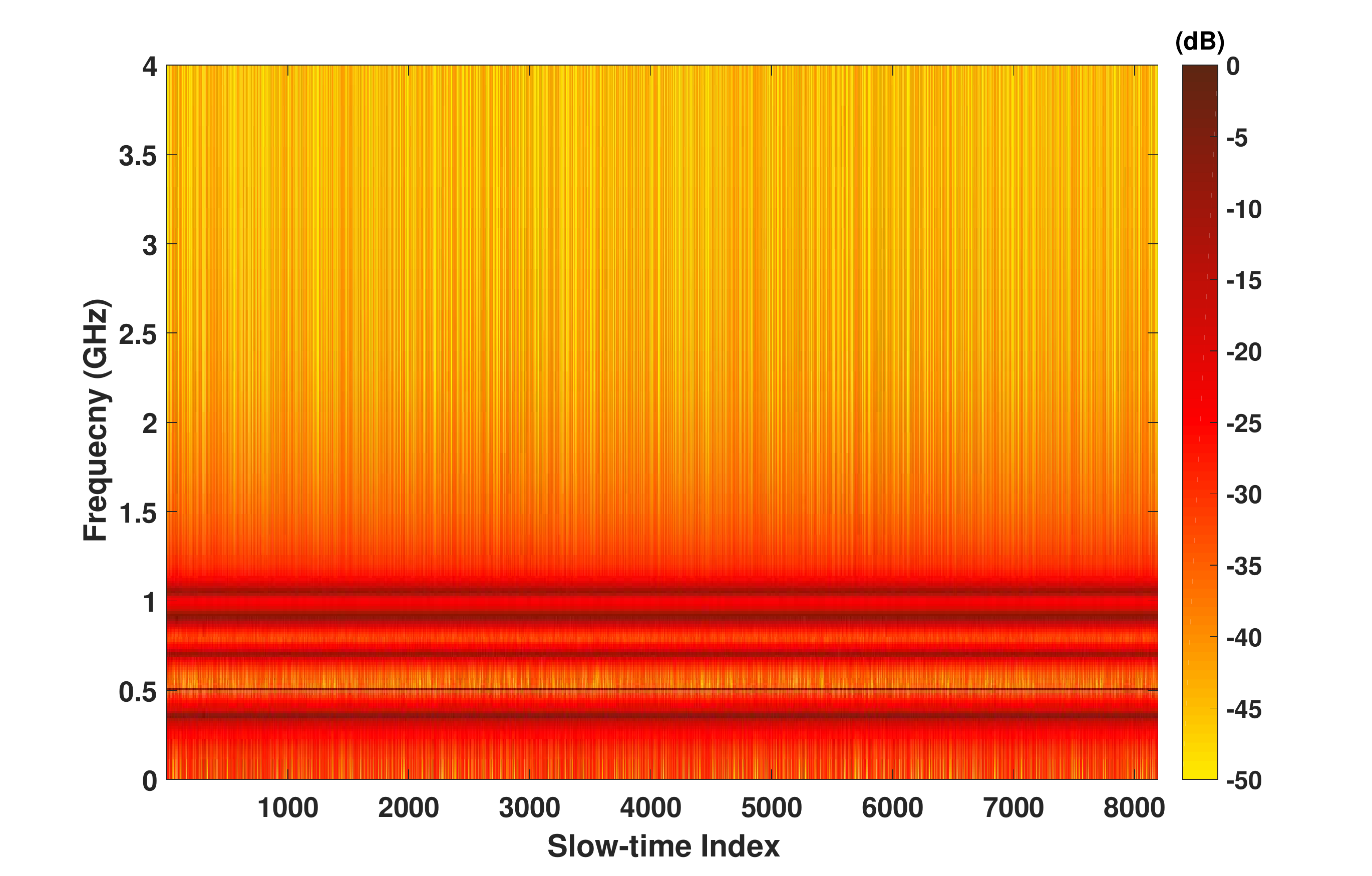}
\caption{Fast-frequency spectrum of the simulated RFI versus slow-time}
\label{fig:spec_RFI_sim}
\end{figure}
We first present the results of the three one-bit weighted SPICE algorithms for the case of simulated RFI data sets. Since the magnitudes of the RFI sources usually do not change greatly during the different PRIs within a CPI, we simulate the RFI sources as a sum of sinusoids with amplitudes and frequencies fixed within a CPI and phases varying randomly and independently with slow-time with a uniform distribution over $[0, 2\pi)$. The $(n,m)$-th element of the simulated RFI ${\bm F}^{\rm sim}$ can be written as follows:
\begin{equation}\begin{split}
f^{\rm sim}_{n,m} &= \sum_{q=1}^Q a_q^{\rm sim} \sin(\omega_q^{\rm sim}(n-1)+\phi_{q,m}^{\rm sim}),\\
\end{split}\end{equation}
where $\frac{a_q^{\rm sim}}{a_1^{\rm sim}}, q=2,\dots,Q$, is the amplitude ratio of the $q$-th RFI source relative to the first one. The parameters of the simulated RFI data are shown in Table \ref{tab:sim_RFI} and the RFI spectrum is displayed in Figure \ref{fig:spec_RFI_sim}. When generating the contaminated data sets with different SINR values, the desired RFI can be obtained by varying $a_1^{\rm sim}$ while fixing $\left\{\frac{a_q^{\rm sim}}{a_1^{\rm sim}}\right\}_{q=2}^Q$.
\begin{table}[htbp]
\caption{Simulated RFI Parameters}\label{tab:sim_RFI}
\centering
\resizebox{0.45\textwidth}{!}{
\begin{tabular}{|c|c|c|c|c|c|}
\hline
RFI Frequencies (MHz)& $500$ & $350$ & $700$ & $900$ & $1050$ \\
\hline
RFI Amplitude Ratios & $1$&$0.95$&$0.8$&$0.87$&$0.9$ \\
\hline
\end{tabular}}
\end{table}

\begin{figure}[htbp]
\centering
\subfloat[]{\includegraphics[width=0.45\textwidth, height = 0.3\textwidth]{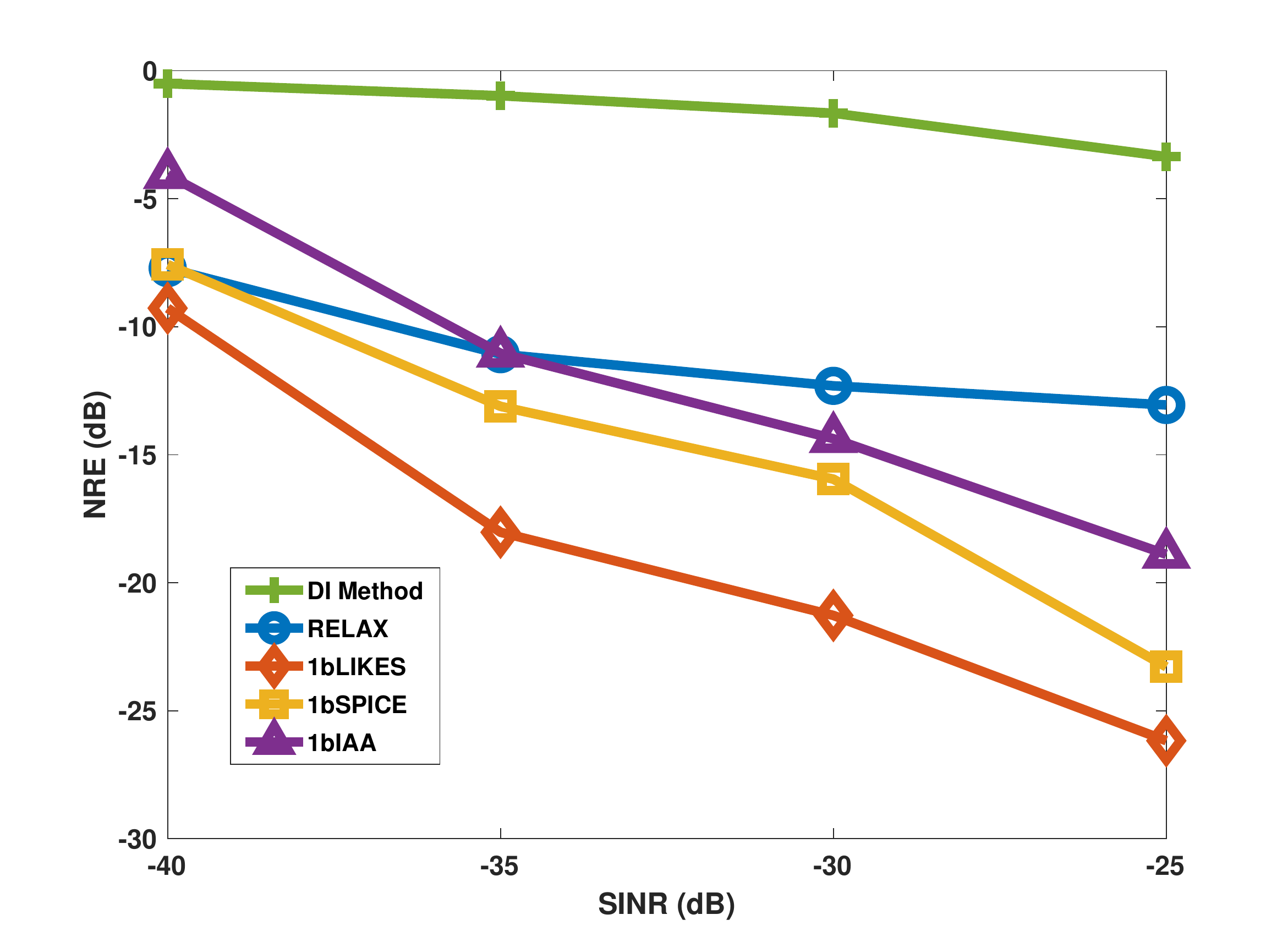}}\\
\subfloat[]{\includegraphics[width=0.45\textwidth, height = 0.3\textwidth]{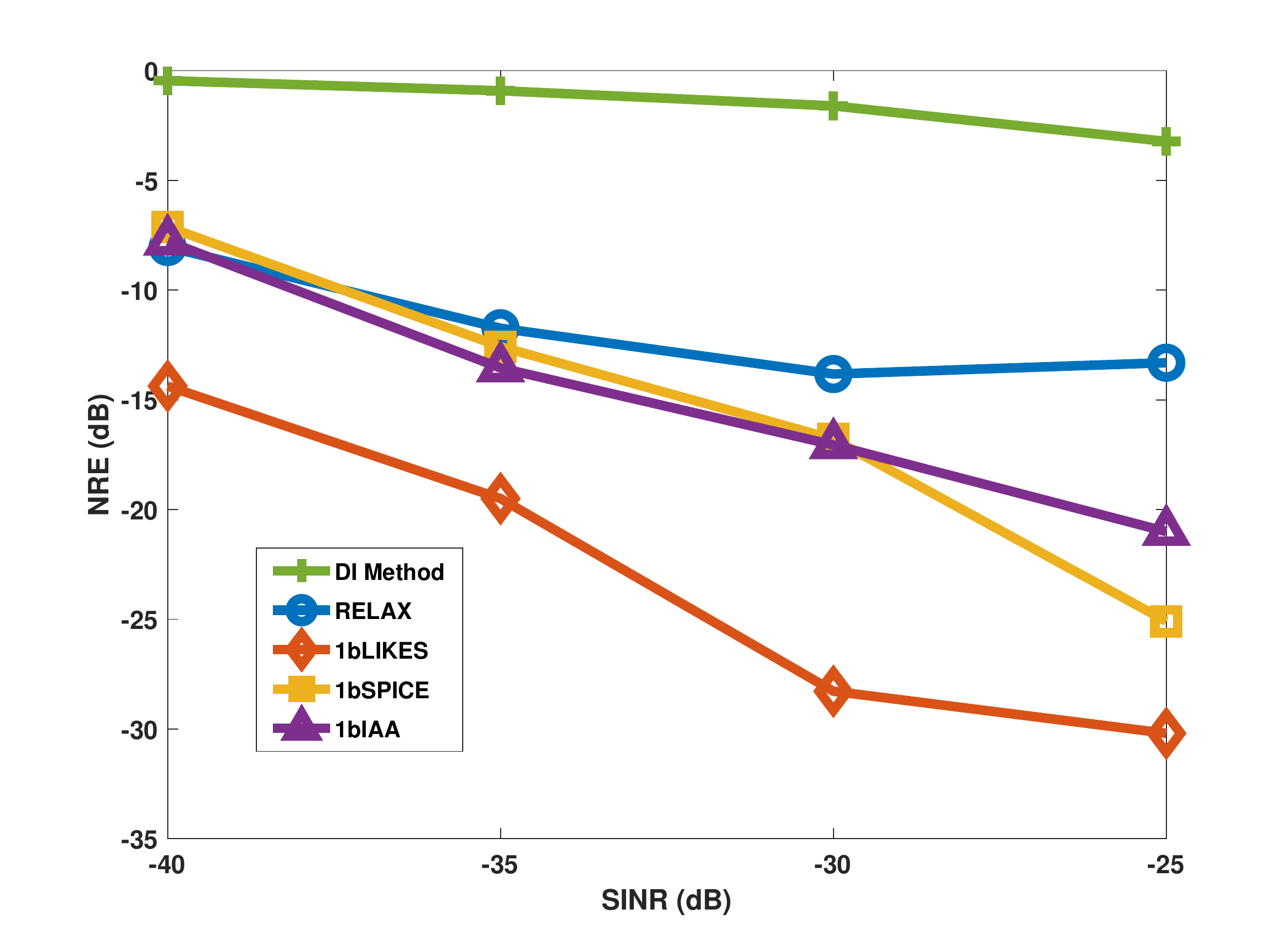}}
\caption{NRE versus SINR results for the three one-bit weighted SPICE algorithms, RELAX, and the DI method for the case of simulated RFI data set. The INR is a) $0$ dB and b) $10$ dB.}
\label{fig:simRFI_NRE}
\end{figure}

\begin{figure*}[htbp]
\centering
\subfloat[]{\includegraphics[width=0.3\textwidth]{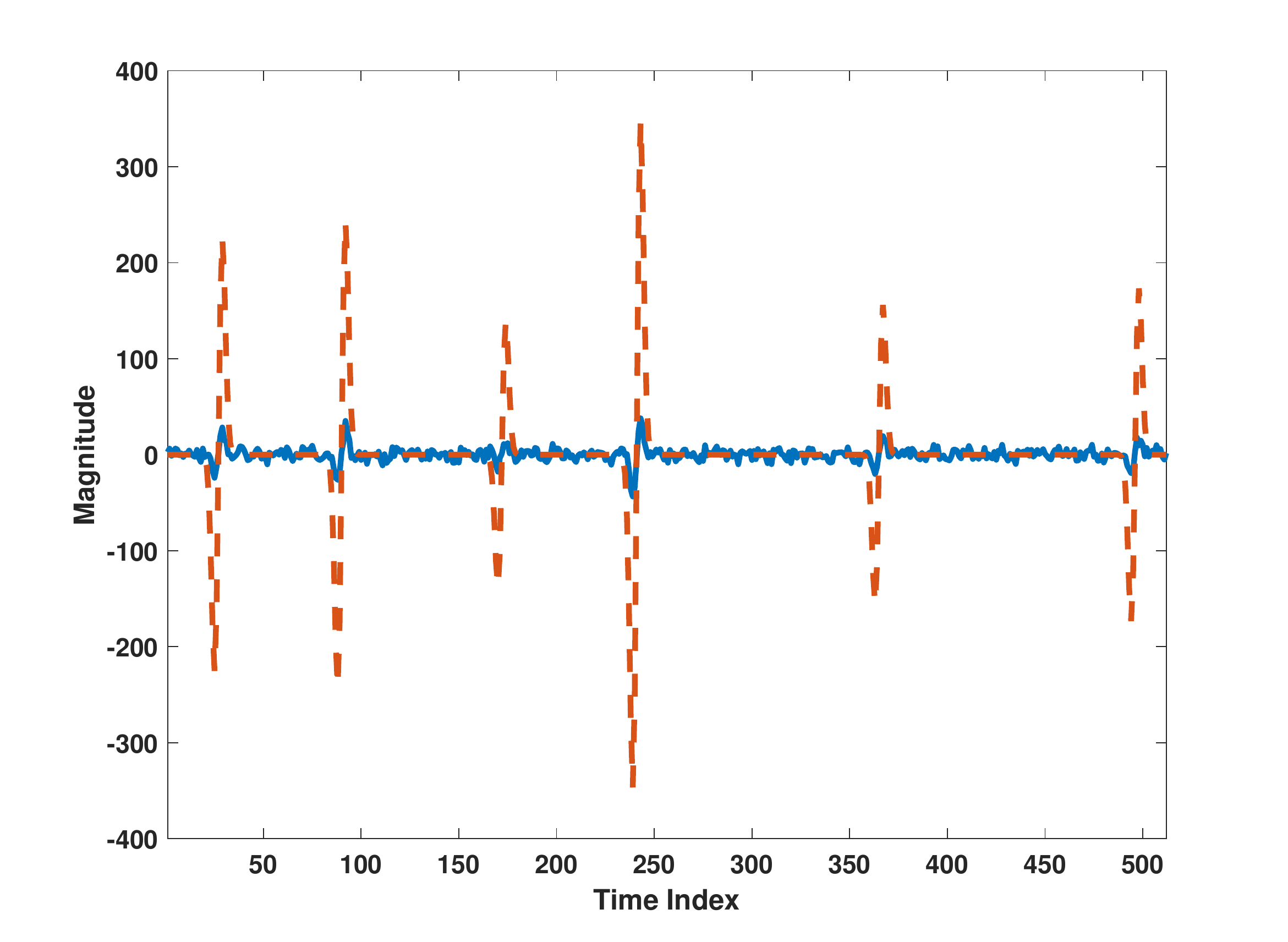}}
\subfloat[]{\includegraphics[width=0.3\textwidth]{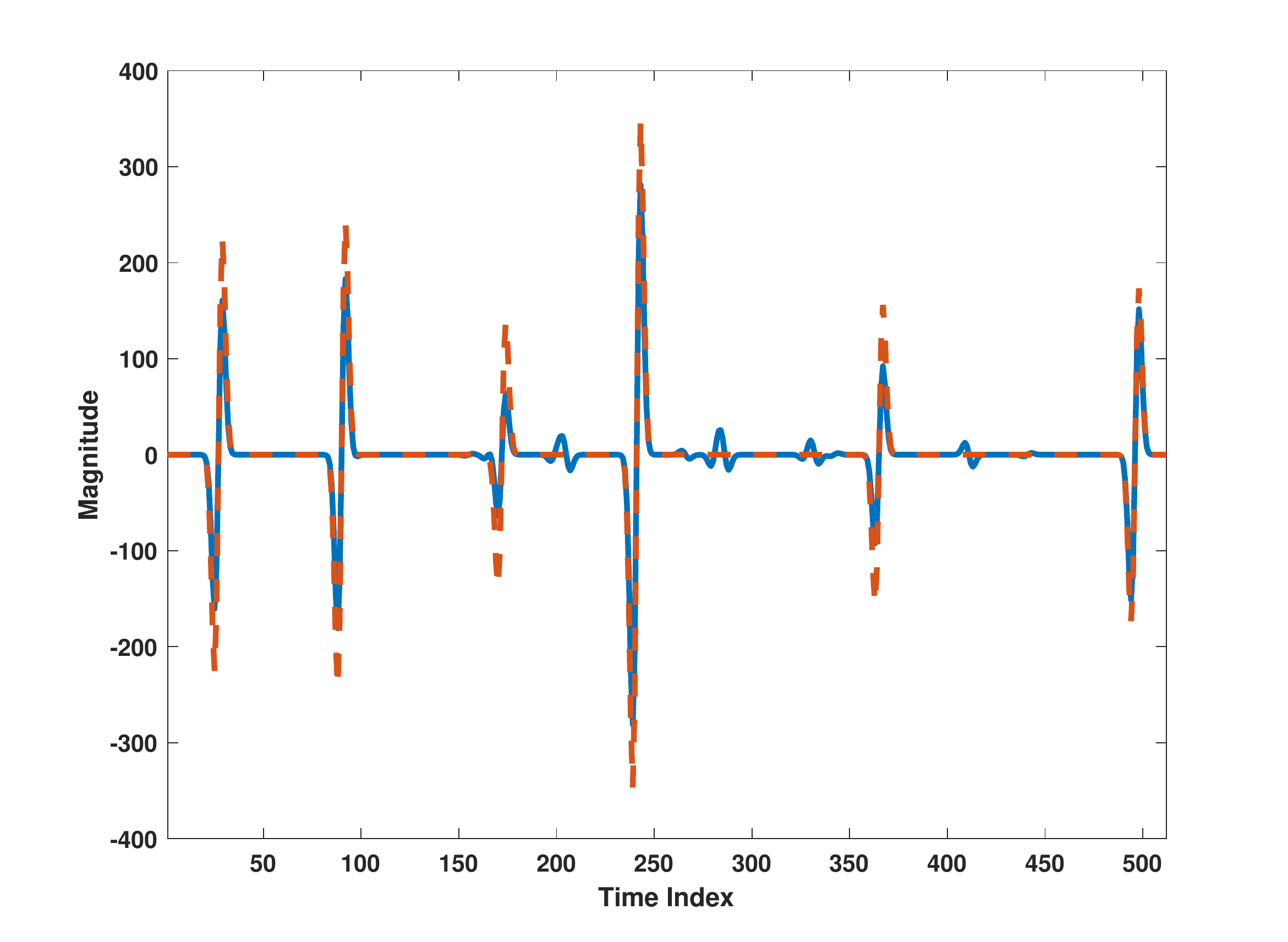}}\\
\subfloat[]{\includegraphics[width=0.3\textwidth]{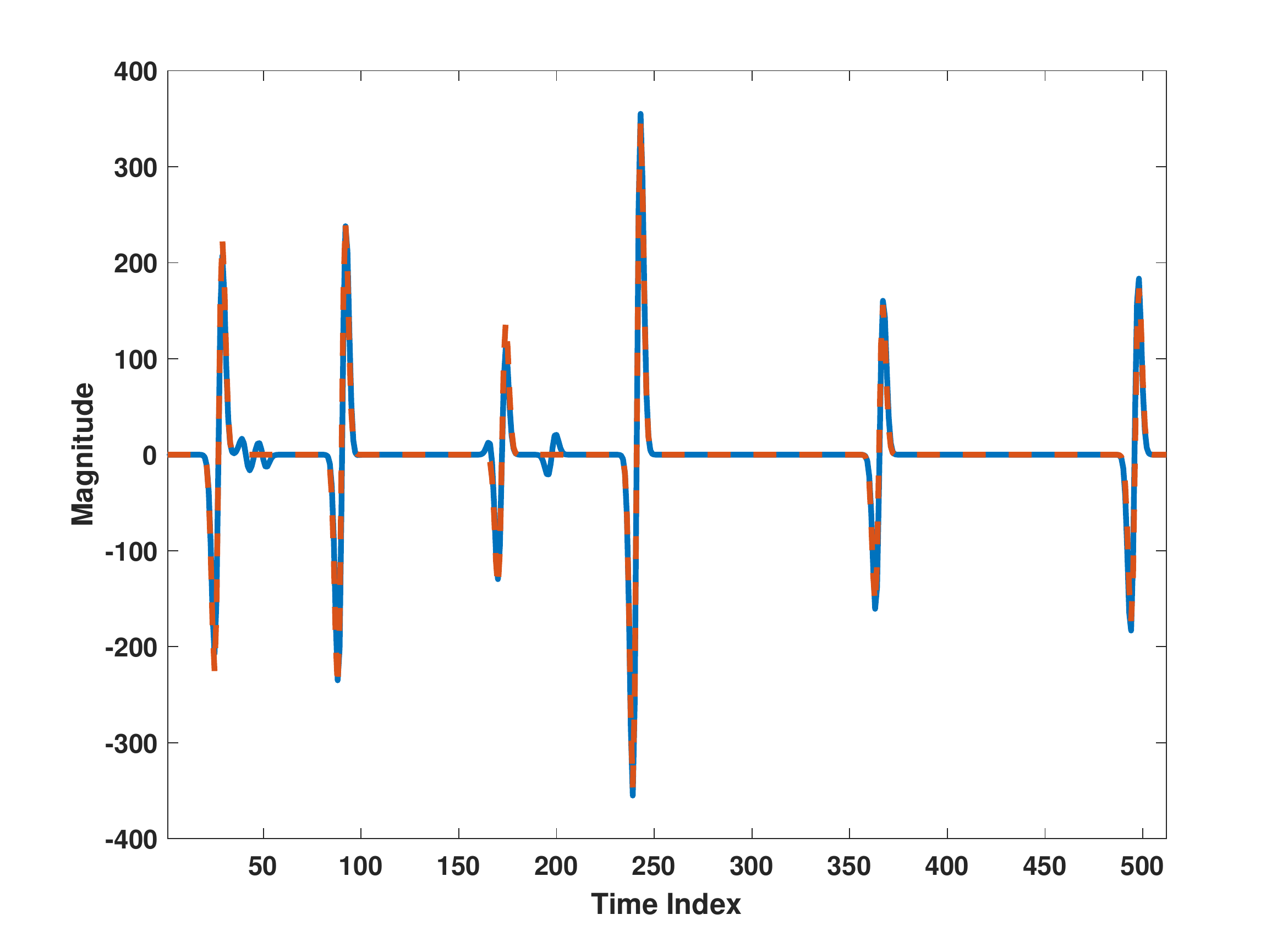}}
\subfloat[]{\includegraphics[width=0.3\textwidth]{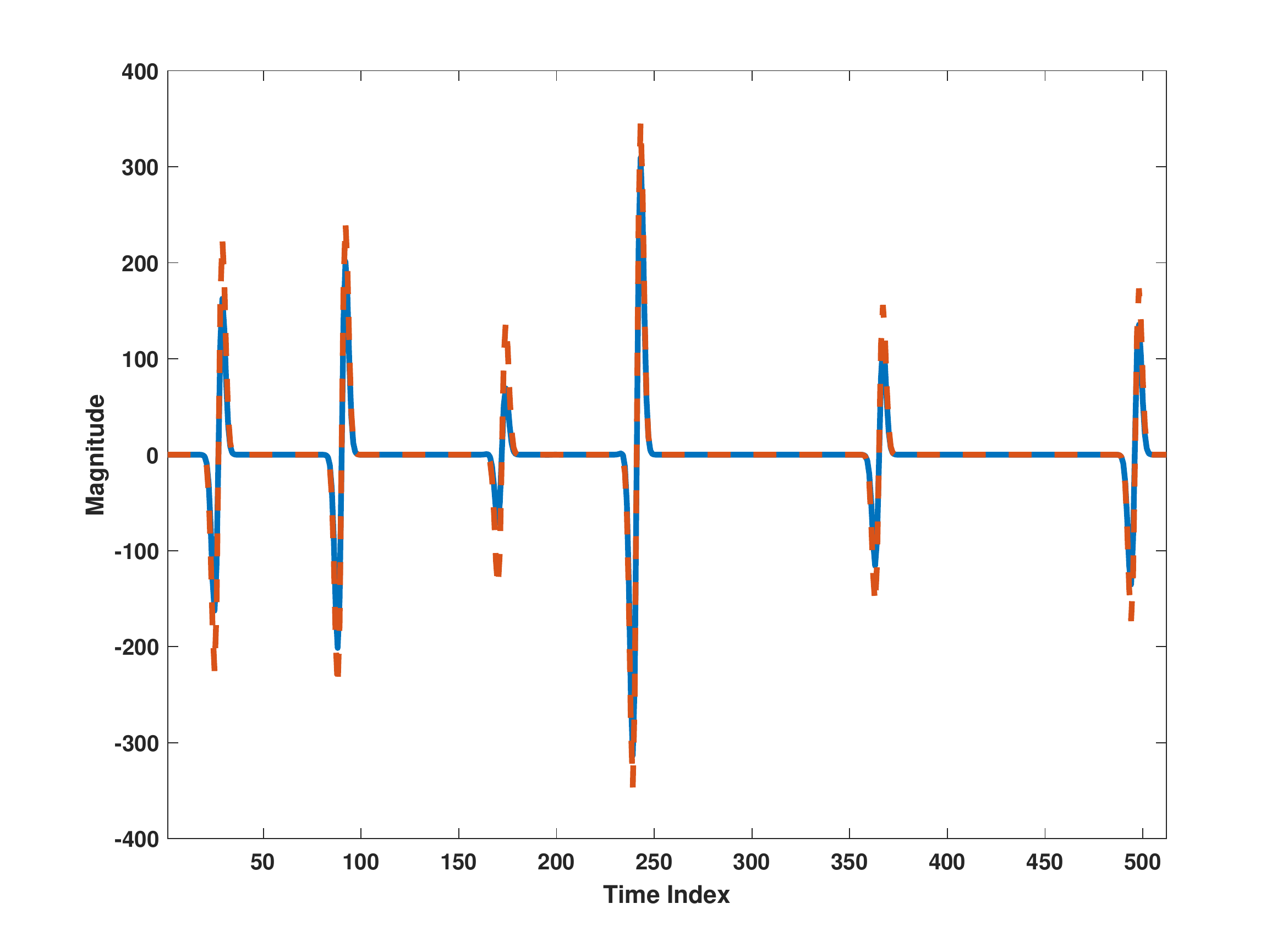}}
\subfloat[]{\includegraphics[width=0.3\textwidth]{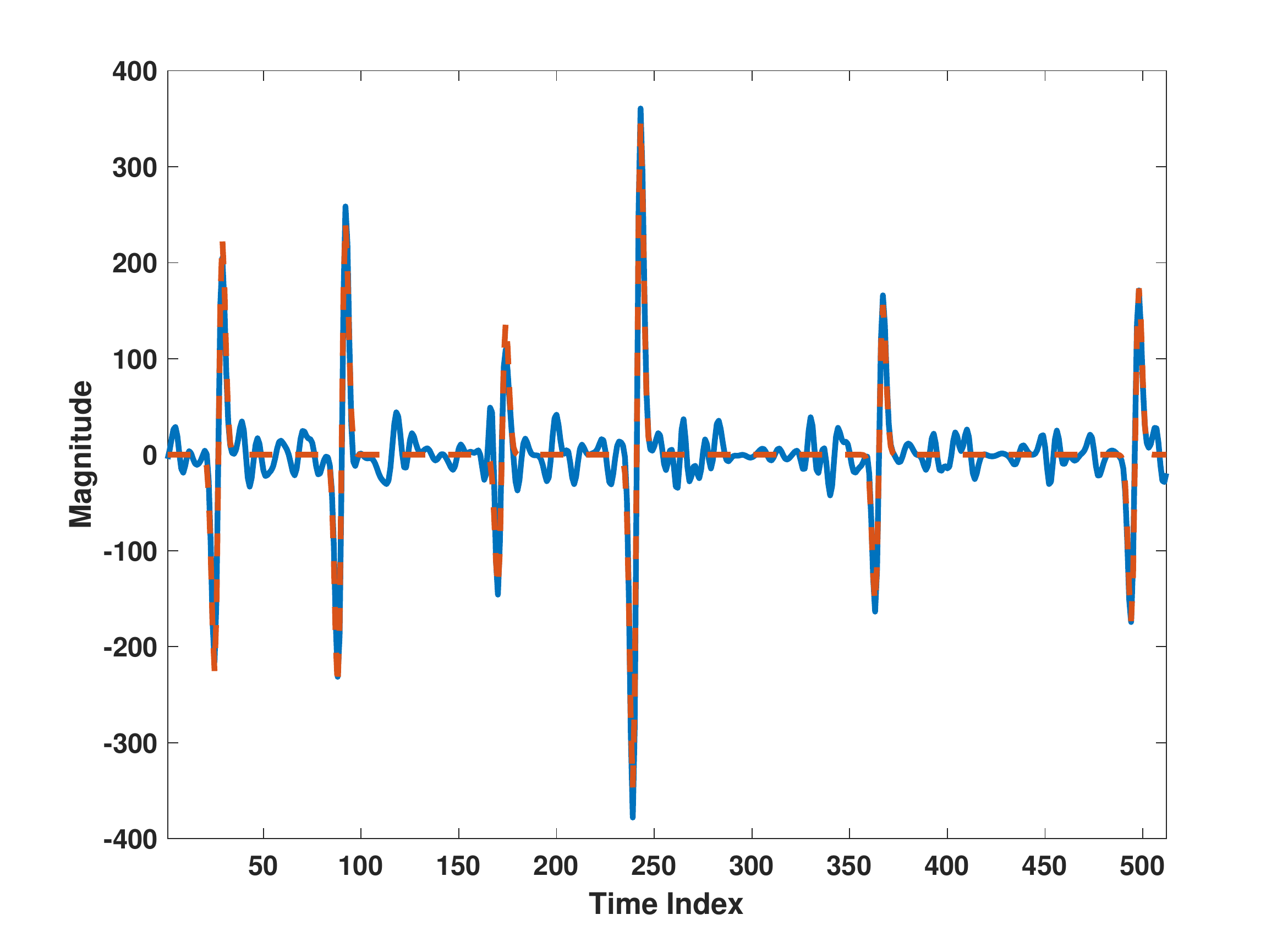}}
\caption{Echo recovery results, for the case of simulated RFI, obtained by using a) the DI method, b) RELAX, c) 1bLIKES, d) 1bSPICE and e) 1bIAA, when SINR = $-35$ dB and INR $=0$ dB.}
\label{fig:simRFI_INR0_35}
\end{figure*}

\begin{figure*}[htbp]
\centering
\subfloat[]{\includegraphics[width=0.3\textwidth]{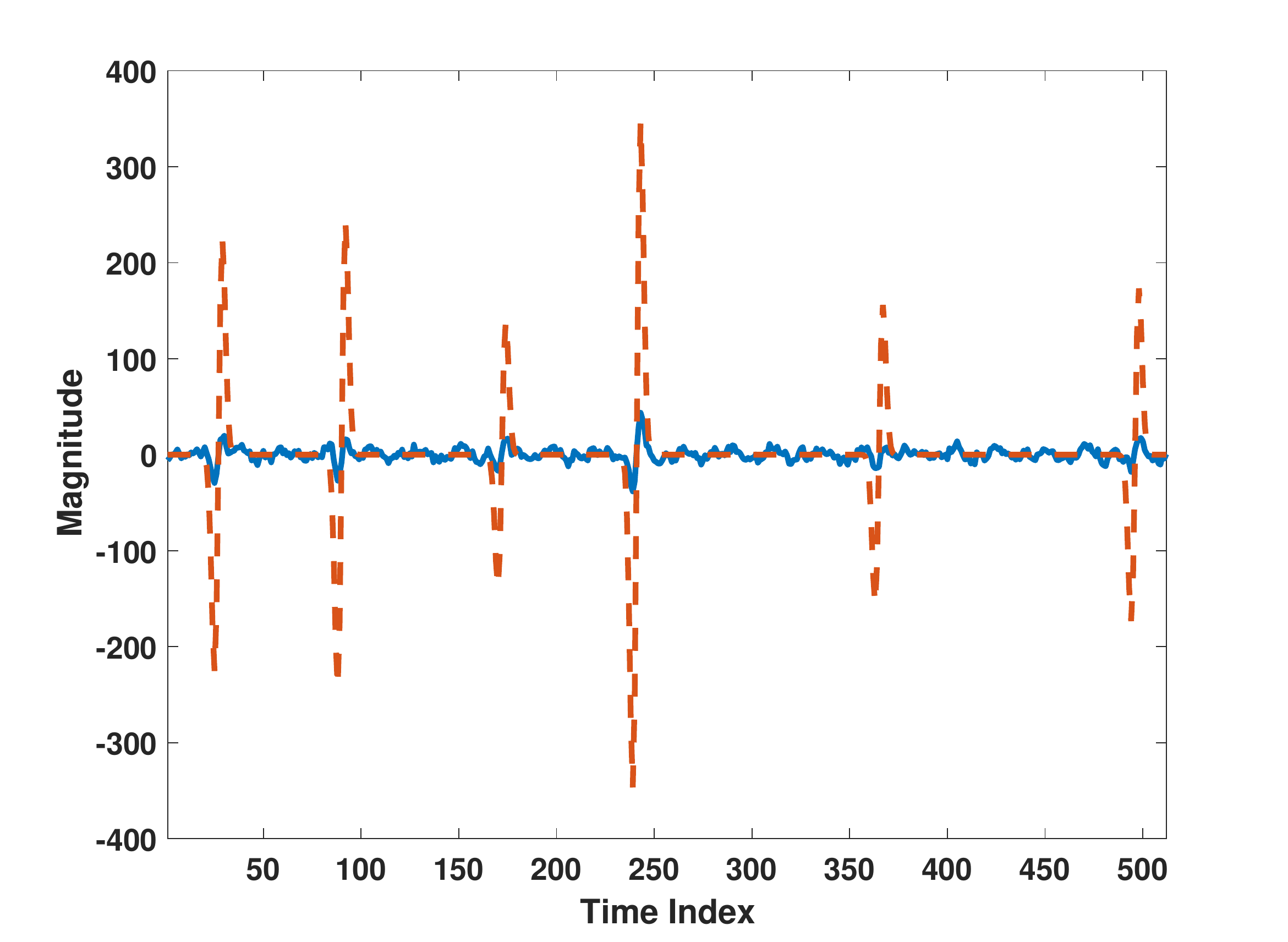}}
\subfloat[]{\includegraphics[width=0.3\textwidth]{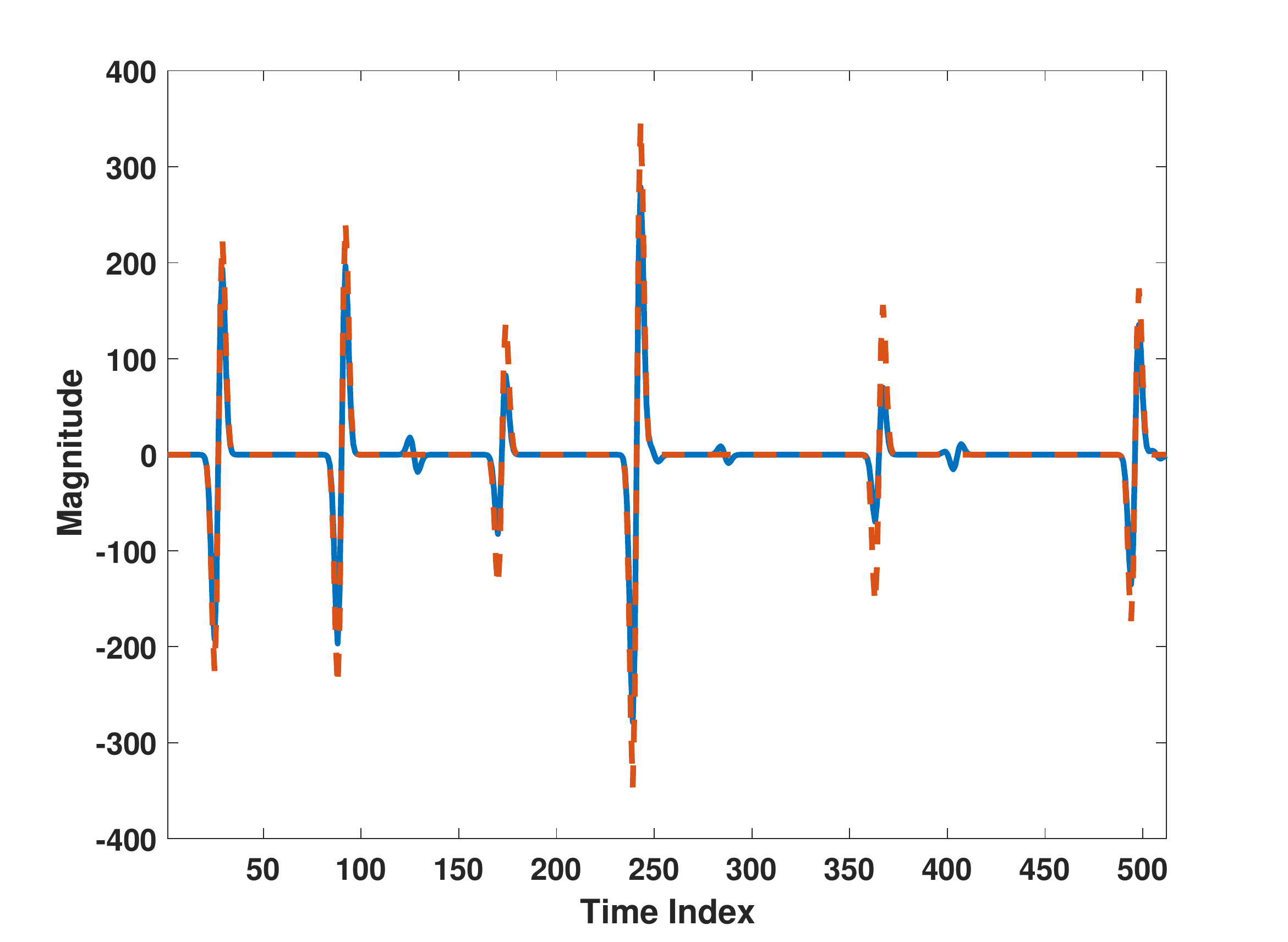}}\\
\subfloat[]{\includegraphics[width=0.3\textwidth]{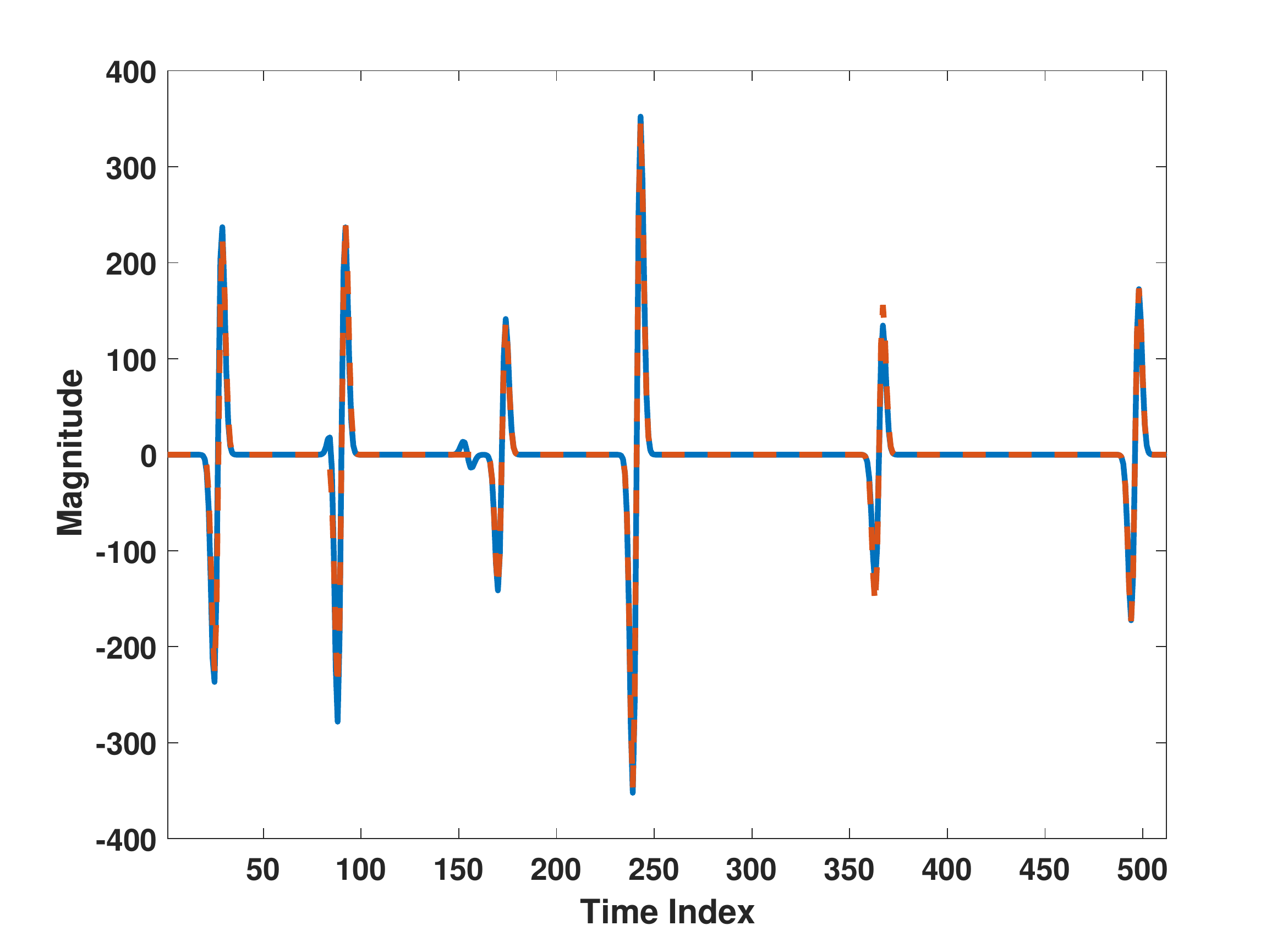}}
\subfloat[]{\includegraphics[width=0.3\textwidth]{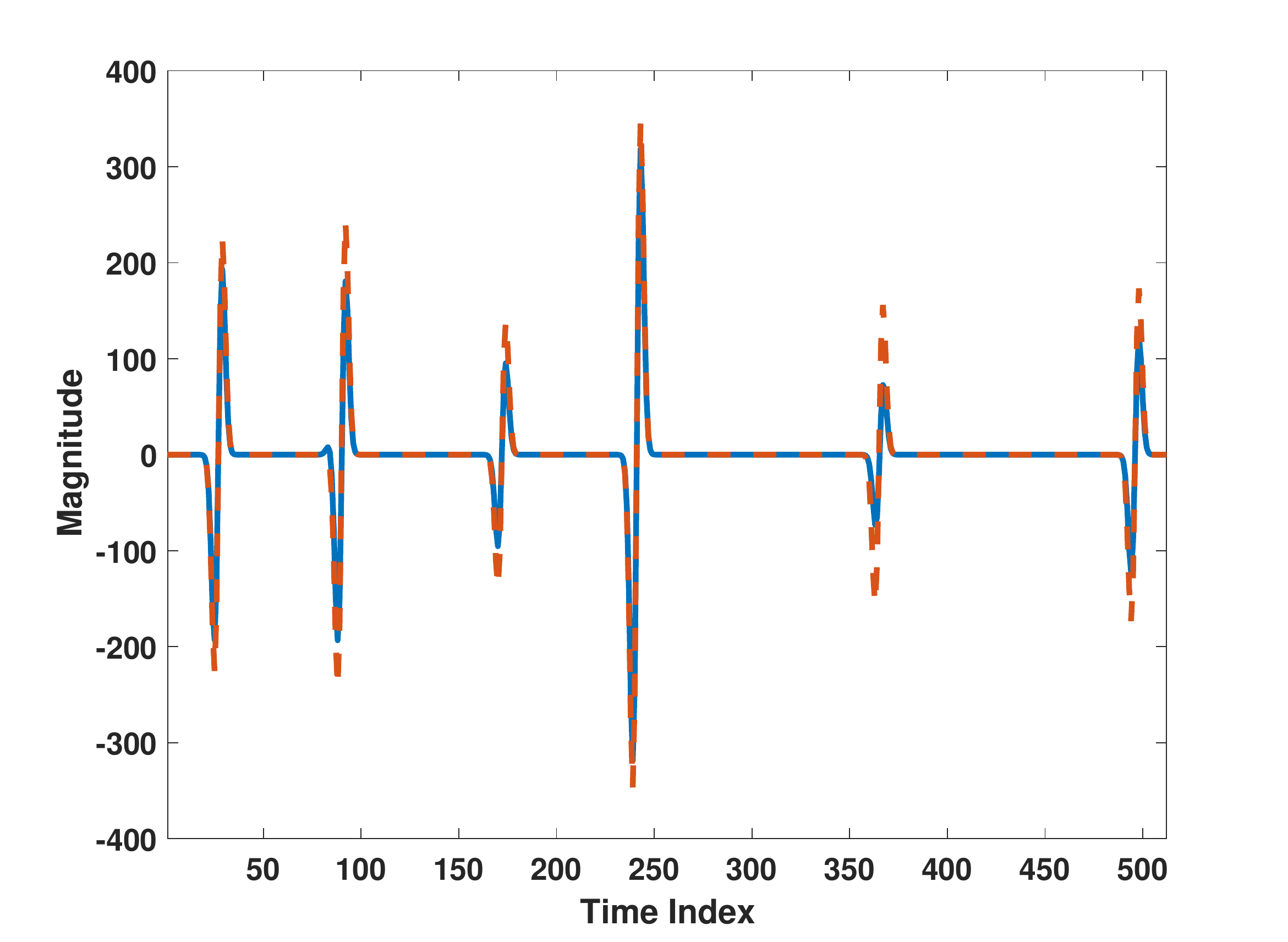}}
\subfloat[]{\includegraphics[width=0.3\textwidth]{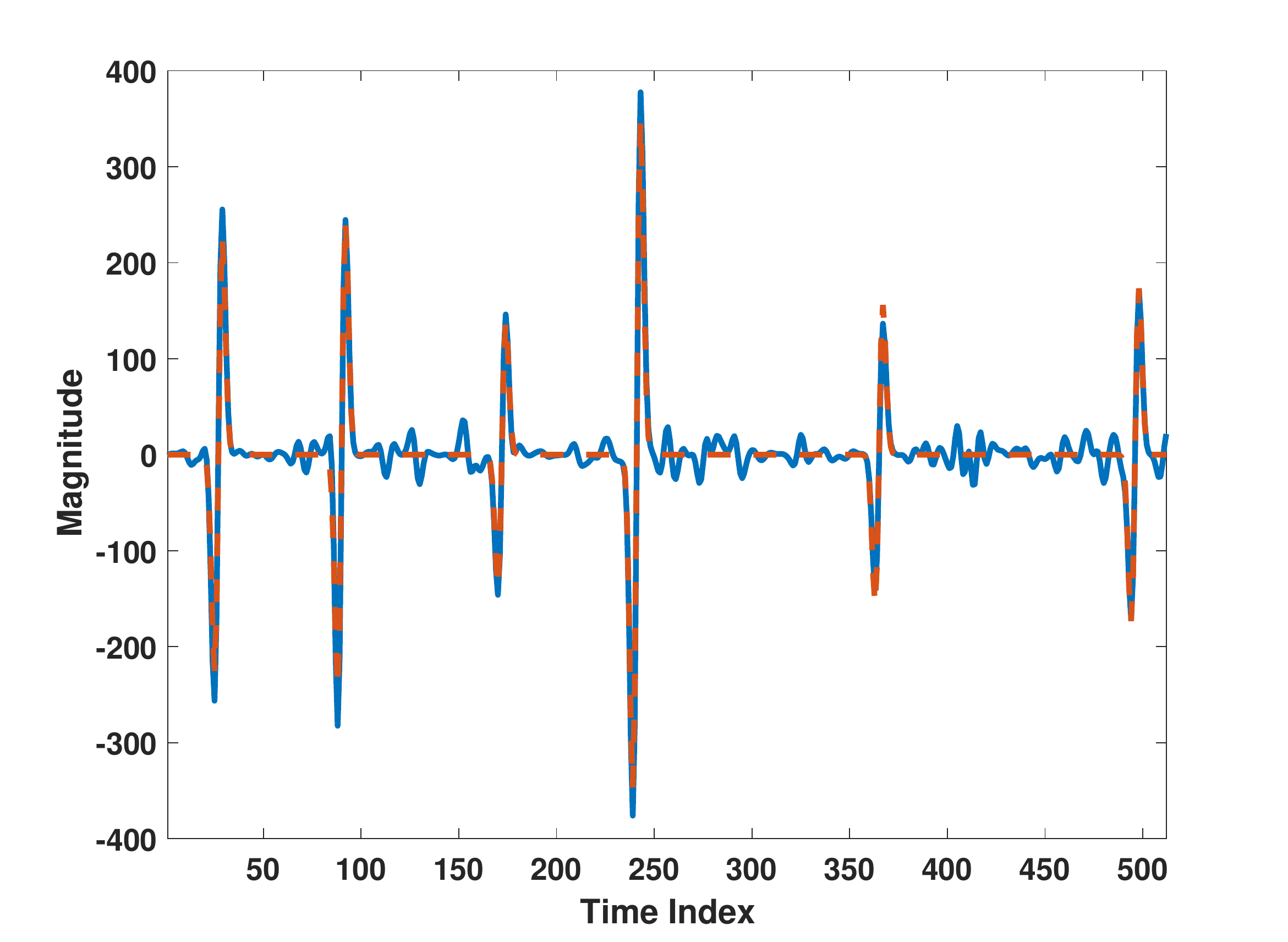}}
\caption{Echo recovery results, for the case of simulated RFI, obtained by using a) the DI method, b) RELAX, c) 1bLIKES, d) 1bSPICE and e) 1bIAA, when SINR = $-35$ dB and INR $=10$ dB.}
\label{fig:simRFI_INR10_35}
\end{figure*}

The NRE versus SINR curves for the case of simulated RFI obtained by the different algorithms are shown in Figures \ref{fig:simRFI_NRE} - \ref{fig:simRFI_INR10_35}.
It is clear that RELAX and the one-bit weighted SPICE algorithms significantly outperform the DI method, for a wide range of SINR values from $-40$ dB to $-25$ dB. Moreover, 1bLIKES provides the best echo recovery performance.

\begin{figure*}[htbp]
\centering
\includegraphics[width=1\textwidth, height=0.3\textwidth]{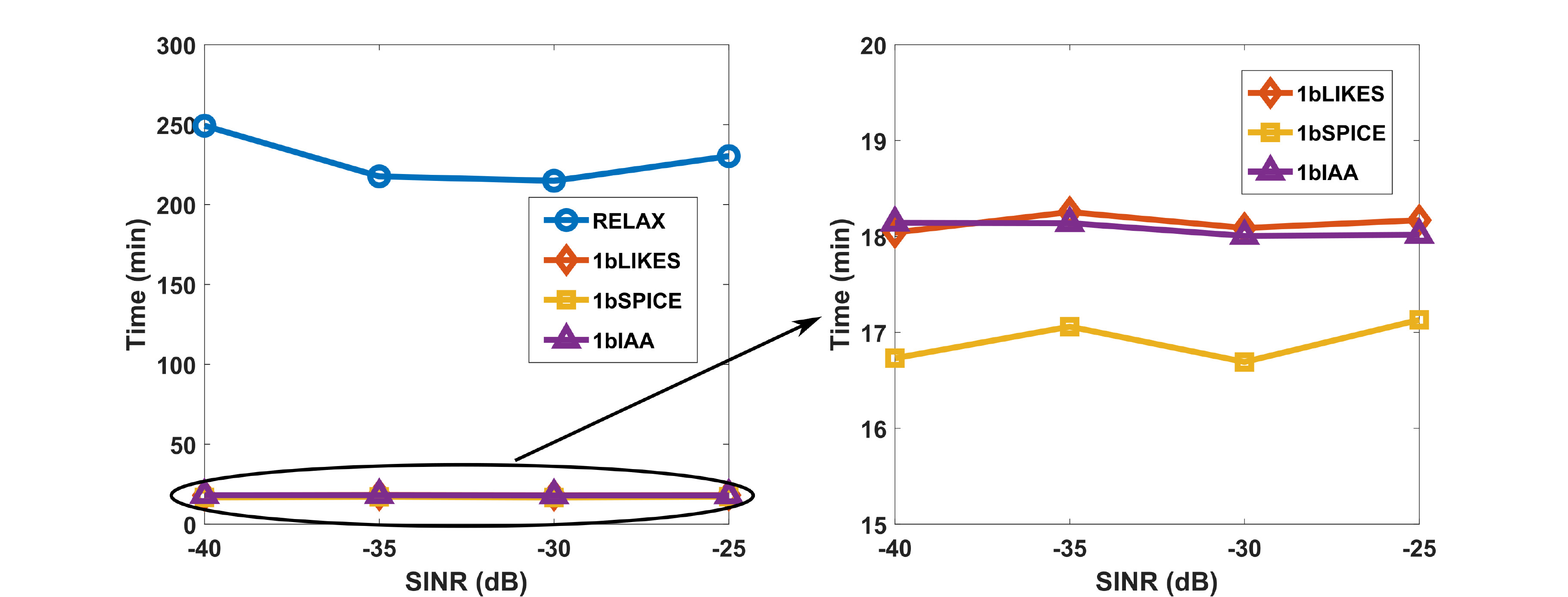}
\caption{Computational times needed by the one-bit weighted SPICE algorithms and RELAX versus SINR for the the case of simulated RFI data sets. The INR is $10$ dB.}
\label{fig:simRFI_time}
\end{figure*}
Figure \ref{fig:simRFI_time} compares the computational times of the three one-bit weighted SPICE algorithms and RELAX as functions of the SINR. The one-bit weighted SPICE algorithms are an order of magnitude faster than the existing RELAX. Note also that 1bSPICE needs slightly less time than 1bLIKES and 1bIAA.
\subsection{Measured RFI case}\label{sec:mea_RFI}
We now shift our attention to the case of measured RFI set collected by the ARL experimental radar receiver \cite{NTD14, NT16}. We use the first $512\times 8192$ samples of the original RFI data set. The spectrum of the measured RFI data set is shown in Figure \ref{fig:spec_RFI_mea}.

\begin{figure*}[htbp]
\centering
\includegraphics[width=0.45\textwidth,, height = 0.3\textwidth]{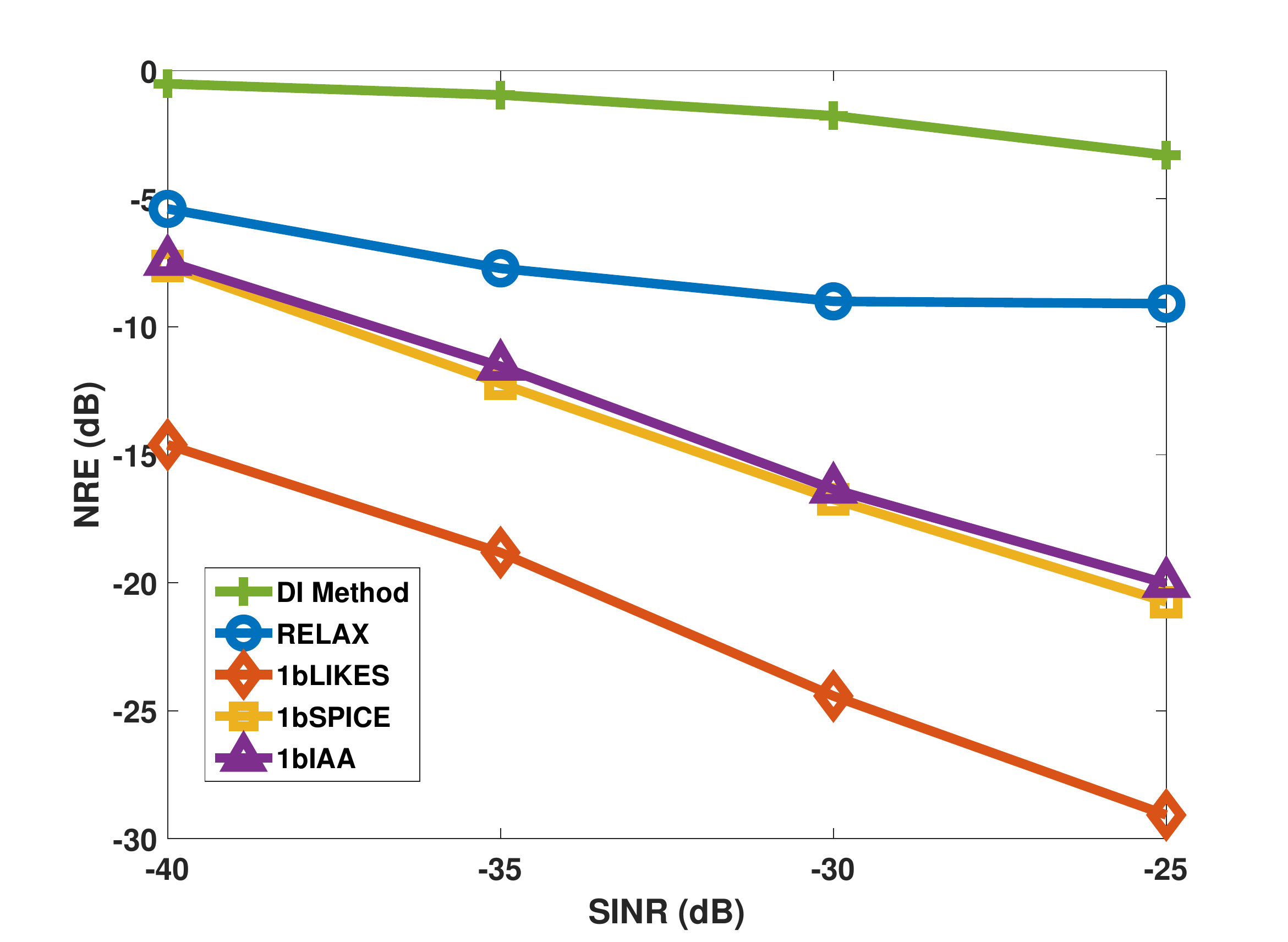}
\caption{NRE versus SINR results for the three one-bit weighted SPICE algorithms, RELAX and the DI method for the case of measured RFI data set.}
\label{fig:meaRFI_NRE}
\end{figure*}

\begin{figure*}[htbp]
\centering
\subfloat[]{\includegraphics[width=0.3\textwidth]{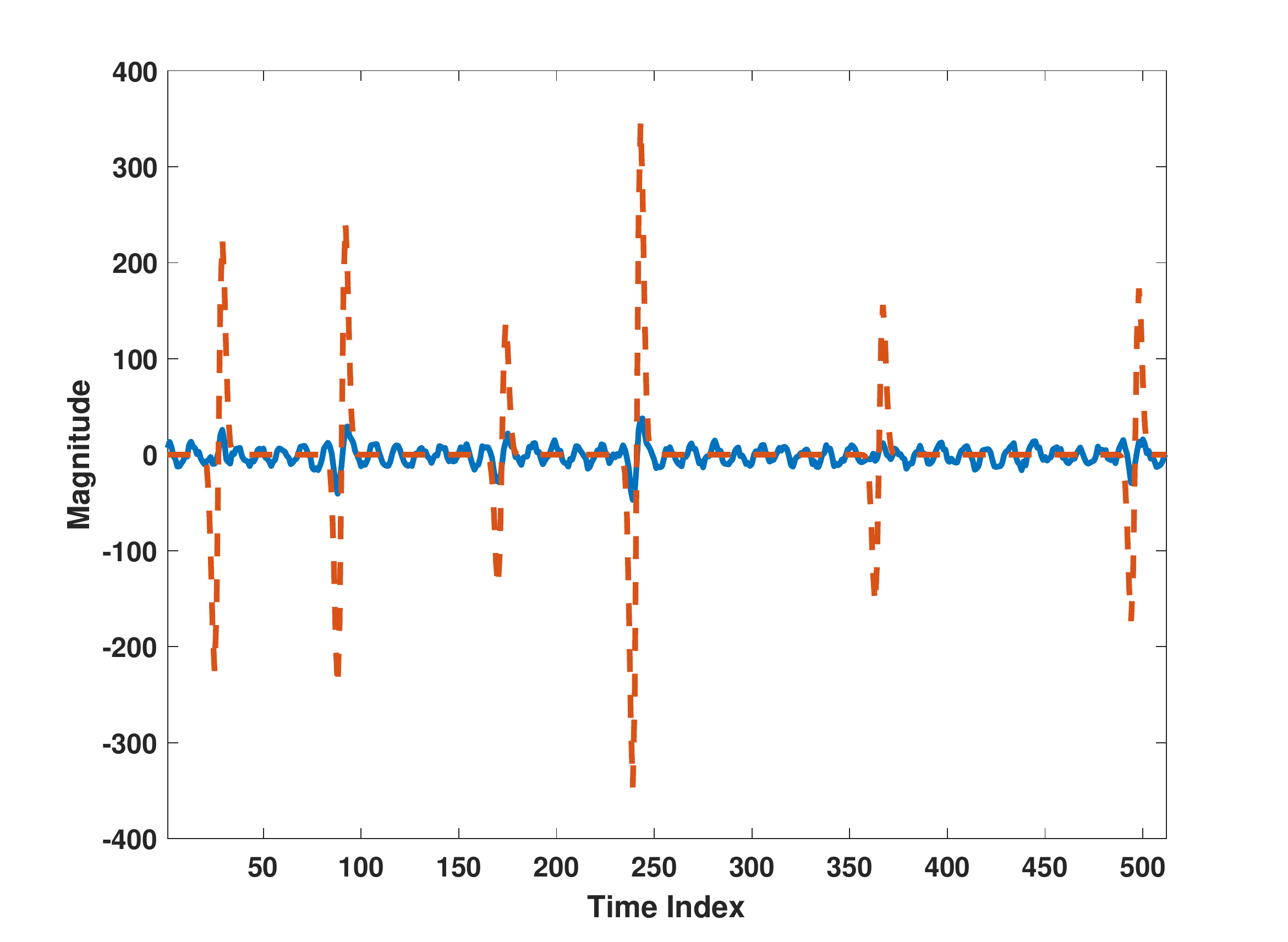}}
\subfloat[]{\includegraphics[width=0.3\textwidth]{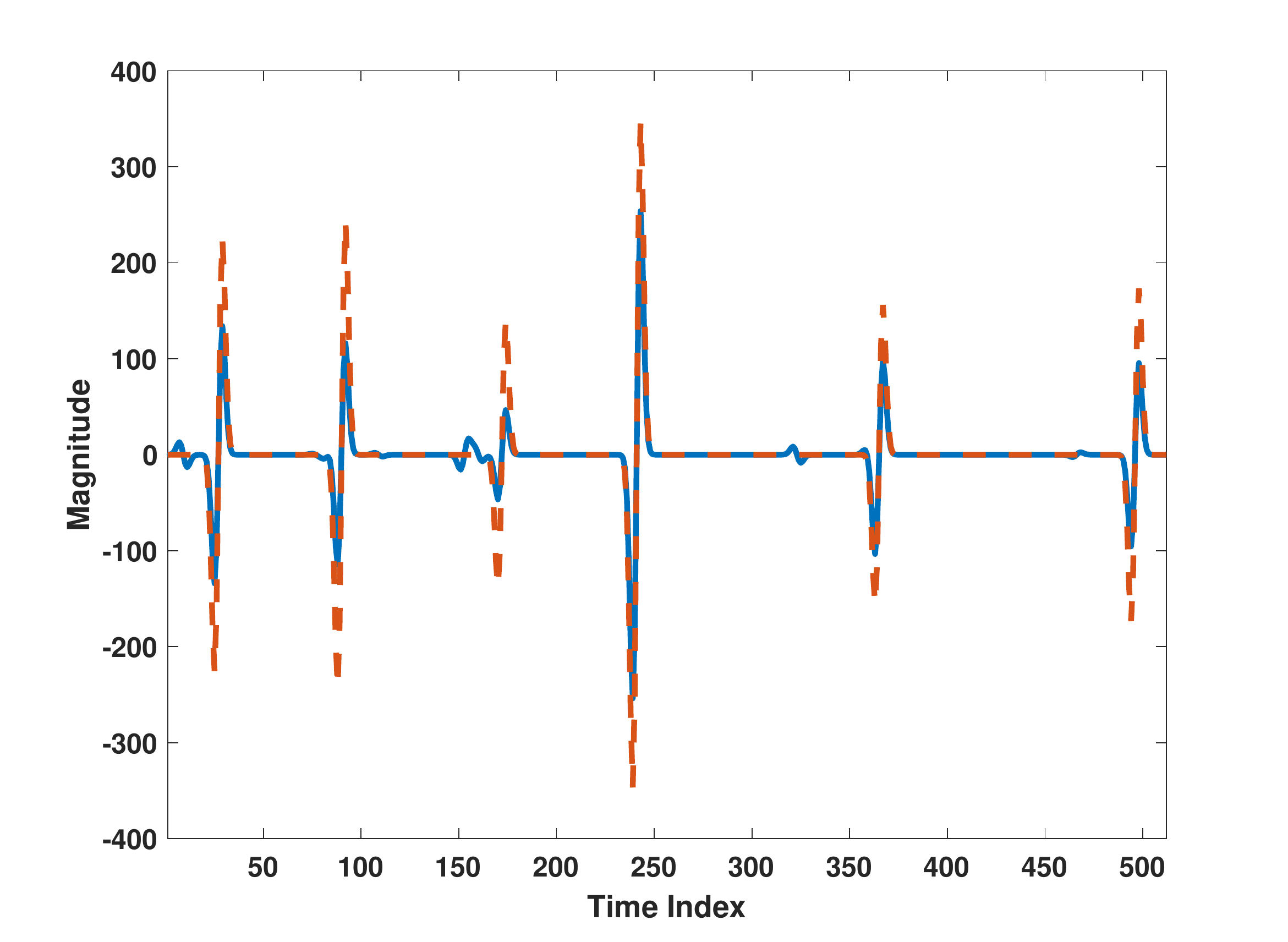}}\\
\subfloat[]{\includegraphics[width=0.3\textwidth]{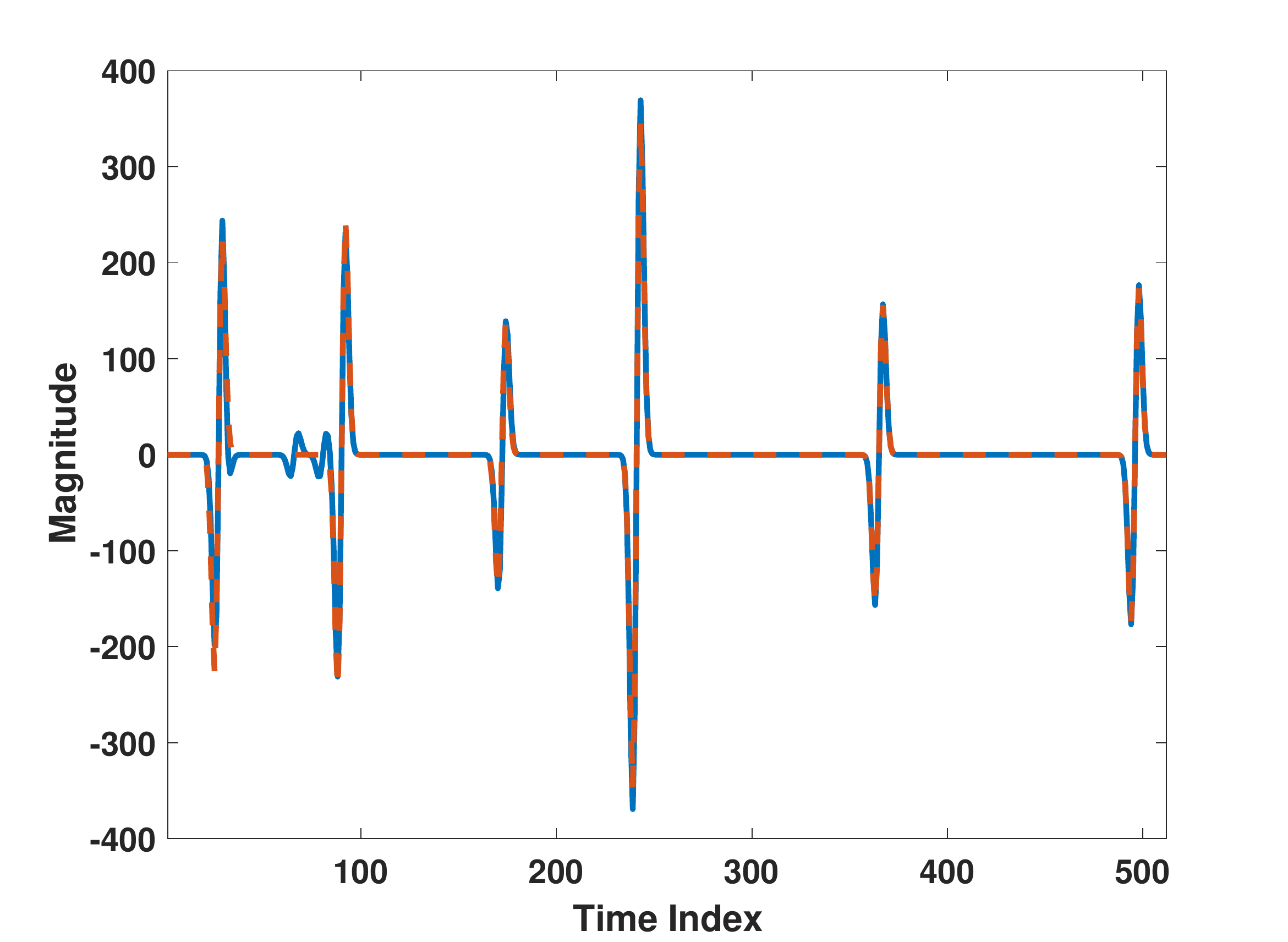}}
\subfloat[]{\includegraphics[width=0.3\textwidth]{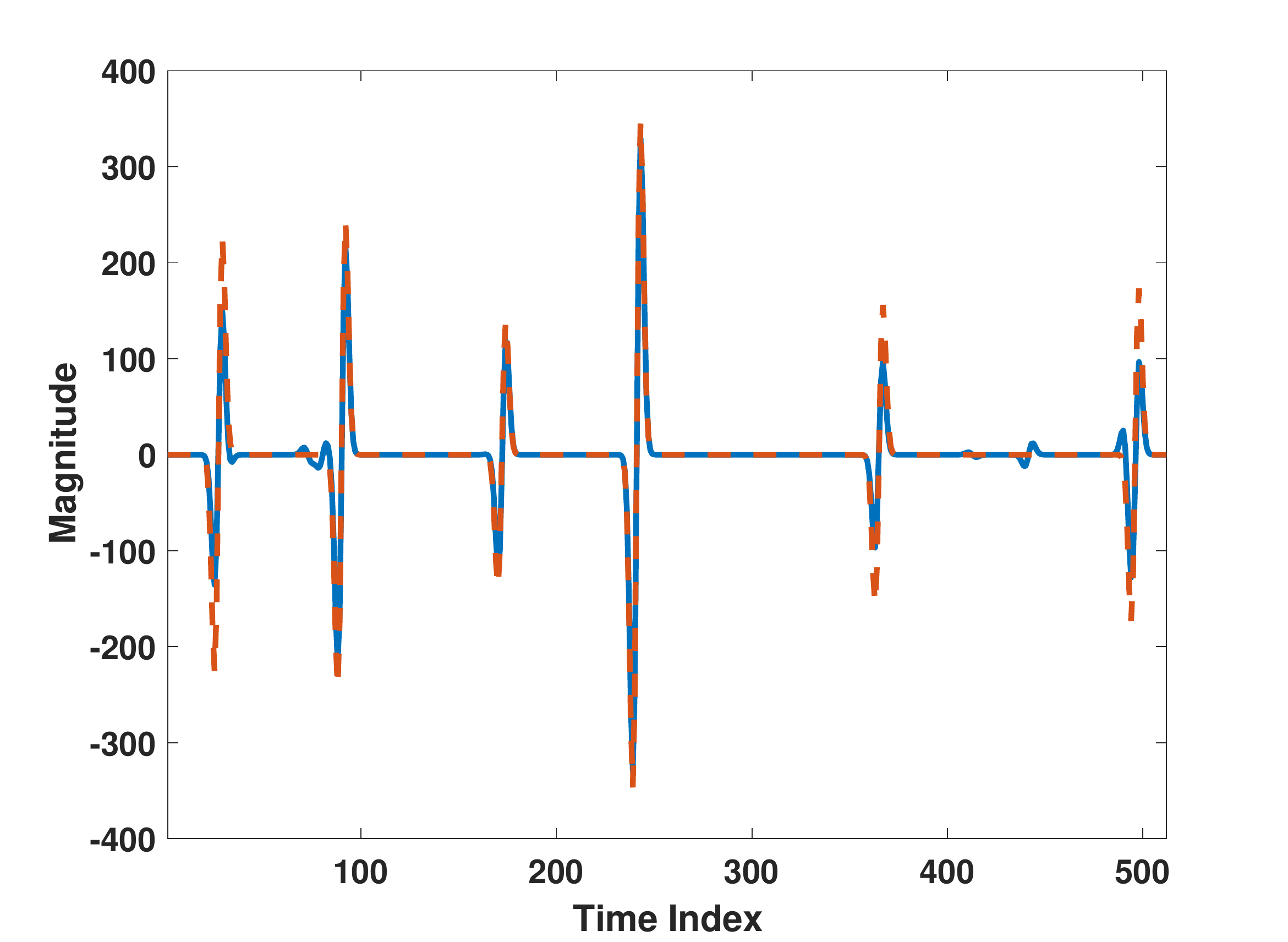}}
\subfloat[]{\includegraphics[width=0.3\textwidth]{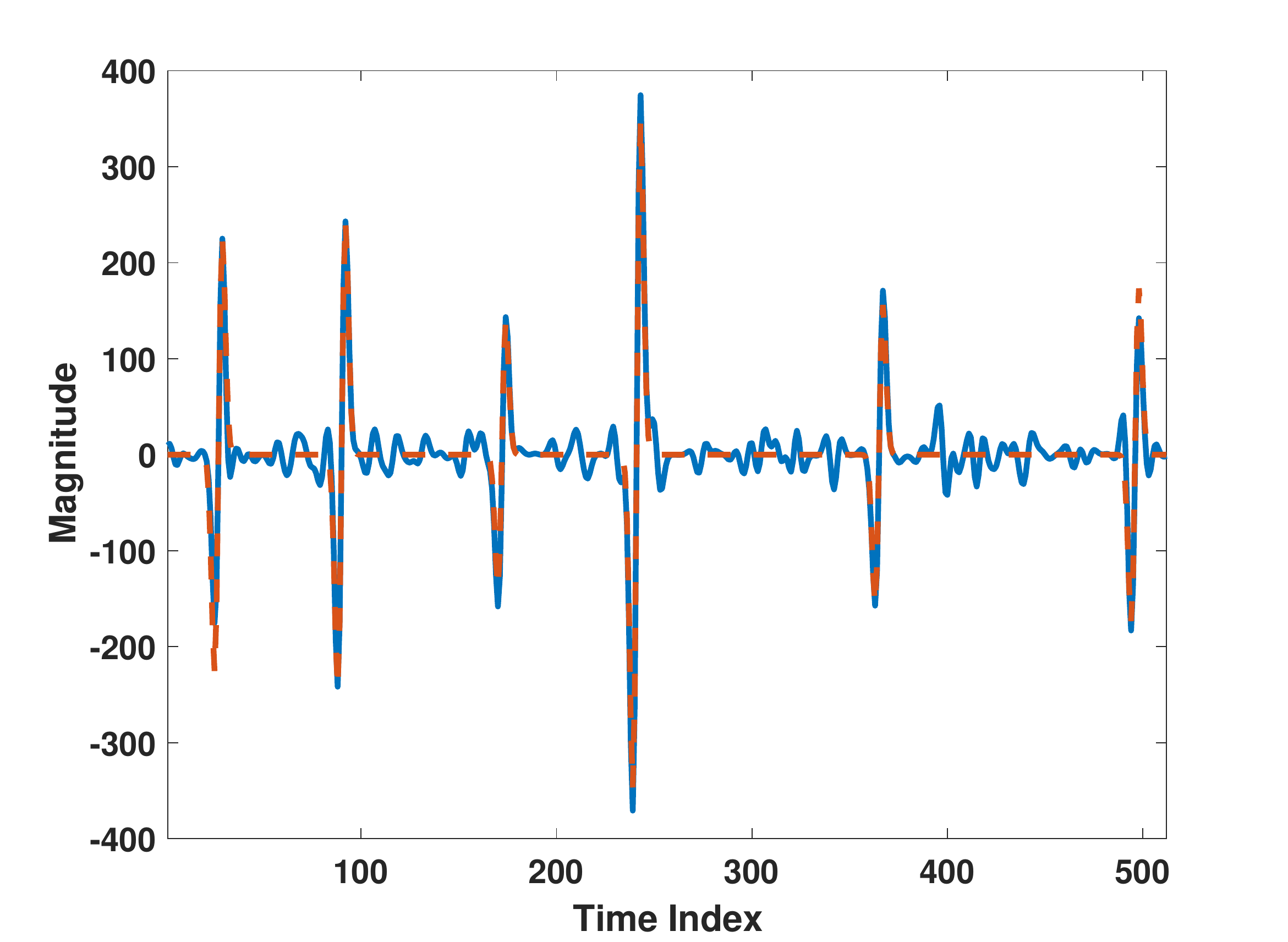}}
\caption{Echo recovery results, for the case of measured RFI, obtained by using a) the DI method, b) RELAX, c) 1bLIKES, d) 1bSPICE and e) 1bIAA, when SINR = $-35$ dB.}
\label{fig:meaRFI_35}
\end{figure*}

The echo recovery results obtained by the different algorithms are shown in Figures \ref{fig:meaRFI_NRE} - \ref{fig:meaRFI_35}. Similar to Section \ref{sec:sim_RFI}, RELAX and the three one-bit weighted SPICE algorithms significantly outperform the DI method, and 1bLIKES provides the best performance.

\begin{figure*}[htbp]
\centering
\includegraphics[width=1\textwidth, height=0.3\textwidth]{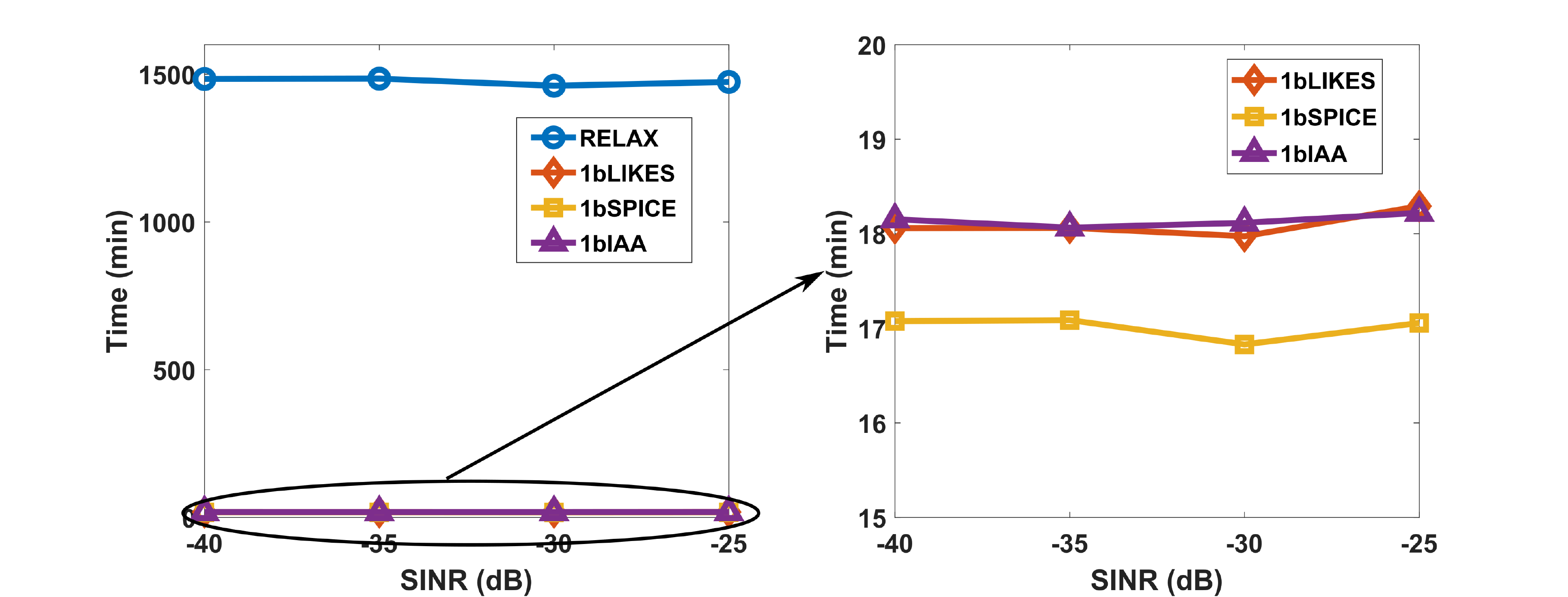}
\caption{Computational times needed by the one-bit weighted SPICE algorithms and RELAX for the case of measured RFI data set and different SINR values. }
\label{fig:meaRFI_time}
\end{figure*}

Figure \ref{fig:meaRFI_time} compares the computational times needed by the one-bit weighted SPICE algorithms and RELAX in this case. The one-bit weighted SPICE algorithms are almost two orders of magnitude faster than RELAX. The number of RFI sources in the measured RFI data set is larger than that in its simulated counterpart. Comparing Figures \ref{fig:meaRFI_time} and \ref{fig:simRFI_time}, it is obvious that when the number of RFI sources increases, the computational burden of RELAX increases sharply since it iteratively estimates the parameters of each RFI source separately. Here too, 1bSPICE consumes slightly less time than 1bLIKES and 1bIAA, due to its simplicity in updating the weight vectors during the iterations.

\section{Conclusions}
We have considered a joint RFI mitigation and sparse echo recovery problem for a one-bit UWB radar system that obtains signed measurements by using the CTBV sampling technique. We have first established a proper data model for the RFI sources and the UWB radar echoes. We then extended a one-bit weighted SPICE framework to jointly mitigate the RFI sources and recover the radar echoes from the signed measurements. Through using different weight vectors in the objective functions, different one-bit weighted SPICE algorithms were obtained which were referred to as 1bLIKES, 1bSPICE and 1bIAA. The one-bit weighted SPICE algorithms jointly estimate all RFI sources and the radar echoes, while the competing technique of \cite{ZRLL20} estimates the parameters of each RFI source separately and iteratively, and recovers the radar echoes after the RFI parameters were estimated. The one-bit weighted SPICE algorithms are computationally much faster that the competing technique. We have provided examples using both simulated and measured RFI data sets to demonstrate that the one-bit weighted SPICE algorithms can significantly outperform the competing technique in regards to the accuracy of the radar echo recovery. The numerical and experimental results also showed that 1bLIKES is the most accurate method of the one-bit weighted SPICE algorithms, whereas 1bSPICE has a slightly lower computational cost than the other two one-bit weighted SPICE algorithms.
% or




\ifCLASSOPTIONcaptionsoff
  \newpage
\fi



%\bibliographystyle{IEEEtran}
\bibliographystyle{IEEEtran}
\bibliography{fast_1bRFI}

\begin{thebibliography}{10}
\providecommand{\url}[1]{#1}
\csname url@samestyle\endcsname
\providecommand{\newblock}{\relax}
\providecommand{\bibinfo}[2]{#2}
\providecommand{\BIBentrySTDinterwordspacing}{\spaceskip=0pt\relax}
\providecommand{\BIBentryALTinterwordstretchfactor}{4}
\providecommand{\BIBentryALTinterwordspacing}{\spaceskip=\fontdimen2\font plus
\BIBentryALTinterwordstretchfactor\fontdimen3\font minus
  \fontdimen4\font\relax}
\providecommand{\BIBforeignlanguage}[2]{{%
\expandafter\ifx\csname l@#1\endcsname\relax
\typeout{** WARNING: IEEEtran.bst: No hyphenation pattern has been}%
\typeout{** loaded for the language `#1'. Using the pattern for}%
\typeout{** the default language instead.}%
\else
\language=\csname l@#1\endcsname
\fi
#2}}
\providecommand{\BIBdecl}{\relax}
\BIBdecl

\bibitem{CHND01}
C.~{Chen}, M.~B. {Higgins}, K.~{O'Neill}, and R.~{Detsch},
  ``Ultrawide-bandwidth fully-polarimetric ground penetrating radar
  classification of subsurface unexploded ordinance,'' \emph{IEEE Transactions
  on Geoscience and Remote Sensing}, vol.~39, no.~6, pp. 1221--1230, June 2001.

\bibitem{XN01}
X.~{Xu} and R.~M. {Narayanan}, ``{FOPEN} {SAR} imaging using {UWB}
  step-frequency and random noise waveforms,'' \emph{IEEE Transactions on
  Aerospace and Electronic Systems}, vol.~37, no.~4, pp. 1287--1300, Oct 2001.

\bibitem{YLML06}
A.~G. {Yarovoy}, L.~P. {Ligthart}, J.~{Matuzas}, and B.~{Levitas}, ``{UWB}
  radar for human being detection,'' \emph{IEEE Aerospace and Electronic
  Systems Magazine}, vol.~21, no.~3, pp. 10--14, March 2006.

\bibitem{SLL20}
X.~{Shang}, J.~{Liu}, and J.~{Li}, ``Multiple object localization and vital
  sign monitoring using {IR-UWB MIMO} radar,'' \emph{IEEE Transactions on
  Aerospace and Electronic Systems}, pp. 1--1, 2020 (Early Access).

\bibitem{ZHLLW18}
B.~{Zhao}, L.~{Huang}, J.~{Li}, M.~{Liu}, and J.~{Wang}, ``Deceptive {SAR}
  jamming based on 1-bit sampling and time-varying thresholds,'' \emph{IEEE
  Journal of Selected Topics in Applied Earth Observations and Remote Sensing},
  vol.~11, no.~3, pp. 939--950, March 2018.

\bibitem{ZHB19}
B.~{Zhao}, L.~{Huang}, and W.~{Bao}, ``One-bit {SAR} imaging based on
  single-frequency thresholds,'' \emph{IEEE Transactions on Geoscience and
  Remote Sensing}, vol.~57, no.~9, pp. 7017--7032, 2019.

\bibitem{Xethru}
\BIBentryALTinterwordspacing
{Novelda AS}. (2017) Xethru: Single-chip radar sensors with sub-mm resolution.
  [Online]. Available: \url{https://www.xethru.com/}
\BIBentrySTDinterwordspacing

\bibitem{HWLLM07}
H.~A. {Hjortland}, D.~T. {Wisland}, T.~S. {Lande}, C.~{Limbodal}, and
  K.~{Meisal}, ``Thresholded samplers for {UWB} impulse radar,'' in \emph{2007
  IEEE International Symposium on Circuits and Systems}, May 2007, pp.
  1210--1213.

\bibitem{KL95}
T.~{Koutsoudis} and L.~A. {Lovas}, ``{RF} interference suppression in
  ultrawideband radar receivers,'' in \emph{Algorithms for Synthetic Aperture
  Radar Imagery II}, vol. 2487.\hskip 1em plus 0.5em minus 0.4em\relax
  International Society for Optics and Photonics, 1995, pp. 107--119.

\bibitem{ZRGL19}
T.~{Zhang}, J.~{Ren}, C.~{Gianelli}, and J.~{Li}, ``{RFI} mitigation for
  one-bit {UWB} radar systems,'' in \emph{53rd Asilomar Conference on Signals,
  Systems, and Computers}, 2019, pp. 1545--1549.

\bibitem{ZRLL20}
\BIBentryALTinterwordspacing
T.~{Zhang}, J.~{Ren}, J.~{Li}, and L.~H. {Nguyen}, ``{RFI} mitigation for
  one-bit {UWB} radar systems,'' arXiv, 2021. [Online]. Available:
  \url{https://arxiv.org/abs/2102.08987}
\BIBentrySTDinterwordspacing

\bibitem{RZLS19}
J.~{Ren}, T.~{Zhang}, J.~{Li}, and P.~{Stoica}, ``Sinusoidal parameter
  estimation from signed measurements via majorization-minimization based
  {RELAX},'' \emph{IEEE Transactions on Signal Processing}, vol.~67, no.~8, pp.
  2173--2186, April 2019.

\bibitem{SZL2014}
P.~{Stoica}, D.~{Zachariah}, and J.~{Li}, ``Weighted {SPICE}: A unifying
  approach for hyperparameter-free sparse estimation,'' \emph{Digital Signal
  Processing}, vol.~33, pp. 1--12, 2014.

\bibitem{SLL20-2}
X.~{Shang}, J.~{Li}, and P.~{Stoica}, ``Weighted {SPICE} algorithms for
  range-{D}oppler imaging using one-bit automotive radar,'' \emph{IEEE Journal
  of Selected Topic in Signal Processing}, 2020 (under review).

\bibitem{SBL11}
P.~{Stoica}, P.~{Babu}, and J.~{Li}, ``New method of sparse parameter
  estimation in separable models and its use for spectral analysis of
  irregularly sampled data,'' \emph{IEEE Transactions on Signal Processing},
  vol.~59, no.~1, pp. 35--47, 2011.

\bibitem{SBL11-2}
------, ``{SPICE}: A sparse covariance-based estimation method for array
  processing,'' \emph{IEEE Transactions on Signal Processing}, vol.~59, no.~2,
  pp. 629--638, 2011.

\bibitem{SB12}
P.~{Stoica} and P.~{Babu}, ``{SPICE} and {LIKES}: Two hyperparameter-free
  methods for sparse-parameter estimation,'' \emph{Signal Processing}, vol.~92,
  no.~7, pp. 1580 -- 1590, 2012.

\bibitem{HL04}
D.~{Hunter} and K.~{Lange}, ``A tutorial on {MM} algorithms,'' \emph{The
  American Statistician}, vol.~58, no.~1, pp. 30--37, 2004.

\bibitem{SS04}
P.~{Stoica} and Y.~{Selen}, ``Cyclic minimizers, majorization techniques, and
  the expectation-maximization algorithm: a refresher,'' \emph{IEEE Signal
  Processing Magazine}, vol.~21, no.~1, pp. 112--114, 2004.

\bibitem{BPCPE11}
S.~{Boyd}, N.~{Parikh}, E.~{Chu}, B.~{Peleato}, and J.~{Eckstein},
  ``Distributed optimization and statistical learning via the alternating
  direction method of multipliers,'' \emph{Foundations and Trends in Machine
  Learning}, vol.~3, no.~1, pp. 1--122, 2011.

\bibitem{NTD14}
L.~H. {Nguyen}, T.~{Tran}, and T.~{Do}, ``Sparse models and sparse recovery for
  ultra-wideband {SAR} applications,'' \emph{IEEE Transactions on Aerospace and
  Electronic Systems}, vol.~50, no.~2, pp. 940--958, April 2014.

\bibitem{NT16}
L.~H. {Nguyen} and T.~D. {Tran}, ``Efficient and robust {RFI} extraction via
  sparse recovery,'' \emph{IEEE Journal of Selected Topics in Applied Earth
  Observations and Remote Sensing}, vol.~9, no.~6, pp. 2104--2117, June 2016.

\bibitem{RNKWS07}
M.~{Ressler}, L.~{Nguyen}, F.~{Koenig}, D.~{Wong}, and G.~{Smith}, ``{The Army
  Research Laboratory ({ARL}) synchronous impulse reconstruction ({SIRE})
  forward-looking radar},'' in \emph{Unmanned Systems Technology IX}, G.~R.
  Gerhart, D.~W. Gage, and C.~M. Shoemaker, Eds., vol. 6561, International
  Society for Optics and Photonics.\hskip 1em plus 0.5em minus 0.4em\relax
  SPIE, 2007, pp. 35 -- 46.

\end{thebibliography}



%\newpage


%\vfill


\end{document}